\documentclass [prb,preprint,superscriptaddress,floatfix,amsmath,amssymb] {revtex4-2}
\usepackage{amsmath}
\usepackage{graphicx}
\usepackage{hyperref}
\usepackage{color}
\usepackage{amssymb}
\usepackage{bm}

\newcommand{\beq} {\begin{equation}}
\newcommand{\eeq} {\end{equation}}
\newcommand{\bea} {\begin{eqnarray}}
\newcommand{\eea} {\end{eqnarray}}
\newcommand{\be} {\begin{equation}}
\newcommand{\ee} {\end{equation}}

\definecolor{darkgreen}{RGB}{0,170,0}

\begin{document}
\title {Free energy and specific heat near a quantum critical point of a metal}
\author{Shang-Shun Zhang}
\affiliation{School of Physics and Astronomy and William I. Fine Theoretical Physics Institute,
University of Minnesota, Minneapolis, MN 55455, USA}
\author{Erez Berg}
\affiliation{Department of Condensed Matter Physics, Weizmann Institute of Sciences, Rehovot, Israel}
\author{Andrey V. Chubukov}
\affiliation{School of Physics and Astronomy and William I. Fine Theoretical Physics Institute,
University of Minnesota, Minneapolis, MN 55455, USA}

\date{\today}
\begin{abstract}
We analyze free energy and specific heat for fermions interacting with gapless bosons at a quantum-critical point (QCP) in a metal. 
We use the Luttinger-Ward-Eliashberg formula for the free energy in the normal state, which includes contributions from bosons, fermions, and their interaction, all expressed via fully dressed fermionic and bosonic propagators. 
The sum of the last two contributions is the free energy $F_\gamma$ of an effective low-energy model of fermions with boson-mediated dynamical 4-fermion interaction $V (\Omega_m) \propto 1/|\Omega_m|^\gamma$ (the $\gamma-$model). 
This purely electronic model has been used to analyze the interplay between non-Fermi liquid (non-FL) behavior and pairing near a QCP, which are both independent of the upper energy cutoff $\Lambda$. 
However, the specific heat $C_\gamma (T)$, obtained from $F_\gamma$, does depend on $\Lambda$. We argue that this dependence is spurious and cancels out, once we include the contribution  from bosons. 
We compare our $C(T)$ with the one obtained within the $\gamma-$model using recently proposed regularization of $F_\gamma$. We argue that for $\gamma <1$, the full $C(T)$ and the regularized $C_\gamma (T)$ differ by a $\gamma-$dependent prefactor, while for $\gamma >1$, the full $C(T)$ is the sum of $C_\gamma (T)$ and the specific heat of free bosons with fully dressed mass. For these $\gamma$, $C_\gamma (T)$ is negative. The authors of Ref. ~\cite{yuz_2} argued that a negative $ C_\gamma (T)$ implies that the normal state becomes unstable at some distance to a QCP. In our calculation, both terms in $C(T)$ come from the same source, and $C_\gamma (T)$ is smaller as long as vertex corrections can be safely neglected. 
We then argue that the normal state remains stable even at a QCP.
 \end{abstract}
\maketitle

\section{ Introduction.}

 In this work we analyze in detail
 the
 free energy and specific heat of a metal
 near a critical point  towards
  a spontaneous
   particle-hole order (Ising-nematic, antiferromagnetic, etc),
   and of an electron-phonon system
    at vanishing  dressed Debye frequency of an optical phonon.
  In all these cases,
  the
  low-energy  physics is described by a  model of
 fermions with Luttinger Fermi surface, coupled by  Yukawa-type interaction to a near-massless boson, which represents either a
  critical fluctuation of a particle-hole order parameter
    or a soft optical phonon~\cite{Nayak1994,Sachdev1995,millis_92,acs,finger_2001,scal_review,2kf2,efetov3,tsvelik,marsiglio2020eliashberg,Chubukov_2020a,Chubukov_2020b,Scalapino2012,Altshuler1994,acf,Oganesyan2001,Metzner2003,rech_2006,*rech_2006_1,efetov,raghu_15,*Wang_H_17,*Wang_H_18,*Fitzpatrick_15,sslee,*sslee2,*lunts_2017}.
  The key motivation for our study is current interest in
    a
    non-Fermi liquid (non-FL)
       behavior near a quantum-critical point (QCP).
     Numerous previous studies have shown
    ~\cite{Altshuler1994,nick_b,acf,Oganesyan2001,Metzner2003,DellAnna2006,*metzner_new,son,son2,rech_2006,metlitski2010quantum2,senthil,
  efetov,cw_14,raghu_15,*Wang_H_17,*Wang_H_18,*Fitzpatrick_15,avi,max_last,sslee,*sslee2,*lunts_2017,
  punk,cm_pom,varma,torroba_1,*torroba_2,Wang2016,review2,
  paper_1,*paper_2,*paper_3,*paper_4,*paper_5,*paper_6,*paper_odd,zhang_22} that at a QCP the self-energy at $T=0$  is
   singular in the frequency domain and  scales as $\Sigma (\omega) \propto \omega^{1-\gamma}$, where
   the
   exponent $\gamma \ll 1$ in weakly anisotropic 3D systems, $\gamma =1/3$ at an Ising-nematic
    and Ising-ferromagnetic
    QCP in 2D, $\gamma \approx 1/2$ at a 2D QCP towards spin or charge density-wave order with a finite momentum, and $\gamma =2$ for an electron-phonon problem.
      It is tempting to associate $1 + d\Sigma/d\omega$ with $m^*/m$ and associate $\omega$ with $T$.  By this reasoning,  the leading term in the specific heat at small $T$,
    $C(T) \propto (m^*/m) T$, should scale  as $T^{1-\gamma}$, i.e. as $T^{2/3}$ at an Ising-nematic QCP, as $T^{1/2}$ at a density-wave QCP, and as $1/T$ for critical electron-phonon problem (although this last behavior obviously cannot extend to $T=0$).
    Our goal is to
    check whether these formulas hold
    in microscopic calculations.

     A more specific
     motivation
     for our work is
     to
     clarify recent studies of
     the free energy for critical fermion-boson systems~\cite{boyack,yuz_1,yuz_2,yuz_3,zhang_22,Erez_c}.  Some of us and others recently analyzed~\cite{paper_1,*paper_2,*paper_3,*paper_4,*paper_5,*paper_6,*paper_odd}
      the interplay between non-FL in the normal state and superconductivity within an effective  low-energy model of
        fermions with
       boson-mediated  dynamical 4-fermion interaction $V(\Omega) \propto 1/|\Omega|^\gamma$ (the $\gamma-$model~\cite{moon_2}). This
        model describes non-FL in the normal state and superconductivity. Both are universal phenomena in the sense that they come from fermions with energies well below the  upper energy cutoff of the model $\Lambda$.  The condensation energy -- the difference between the free energy of a superconductor and of a would be normal state at the same $T$ , is also independent on $\Lambda$ (Ref. \cite{paper_1,*paper_2,*paper_3,*paper_4,*paper_5,*paper_6,*paper_odd}).
      However,
       the free energy of the $\gamma-$model in the normal state is  non-universal, even if we subtract its value at $T=0$.
         Namely, its leading $T-$dependent term scales as $\Lambda T^{1-\gamma}$ (Ref. \cite{zhang_22}).   The corresponding specific heat is then $C(T) \propto \Lambda/T^\gamma$, in variation with the estimate based on the self-energy.
     The authors of Refs. \cite{yuz_2,yuz_3} argued that the dependence of the free energy and the specific heat on $\Lambda$  is
     spurious
      and has to be regularized by  adding the
      counter term
      to the free energy~\cite{yuz_2,yuz_3}, which cancels out
      $\Lambda$ dependence.   Once this is done,
      the regularized specific heat
      becomes independent on $\Lambda$ and scales as $T^{1-\gamma}$,
      as expected
      based on
      the self-energy. However, the regularization comes with the cost:
        the prefactor in $C(T) \propto T^{1-\gamma}$ turns out to be negative for
        $\gamma \ge 1$.
    \footnote{The authors of~\cite{yuz_2,yuz_3} found the prefactor to be negative for $\gamma \geq 2$, which they only considered.
     The  authors of~\cite{zhang_22}
       argued that the prefactor is negative for $\gamma >1$.}

        Taken at a face value, a negative $C(T)$ would imply that the system becomes unstable below a certain $T_{cr}$, when a negative $T^{1-\gamma}$ term, coming from fermion-boson interaction, exceeds  a positive $O(T)$ contribution to $C(T)$
        from free fermions. A potential resolution would be that this instability is preempted by superconductivity,
         but  it turns out that $T_{cr} > T_c$  (Refs \cite{yuz_2,yuz_3,zhang_22}).

    In this work, we analyze free energy and specific heat within the full fermion-boson model
     using the Luttinger-Ward-Eliashberg formula~\cite{lw,Eliashberg} for the variational free energy in the normal state.  We
         assume that
         superconductivity
        is suppressed,
        and
        extend
        the
        normal state analysis down to small $T$.
       Luttinger and Ward argued~\cite{lw}  that the free energy of a system of fermions with
       4-fermion
       interaction
        can be expressed diagrammatically by collecting skeleton diagrams with fully dressed fermionic propagators
        and using conventional rules of the diagrammatic technique, but one has to add to free energy the
        term $F_{el}$, which explicitly contains fermionic self-energy $\Sigma$ (see below).  This additional term is constructed
         such that the stationary condition $\delta F/\delta \Sigma =0$  reproduces  the diagrammatic series for the self-energy.  Eliashberg extended  Luttinger-Ward approach to the case of electron-phonon interaction. He argued that the  free energy for such a system
          is obtained by
          collecting skeleton diagrams with fully dressed fermionic  and bosonic propagators,
           and contains a second
           extra term $F_{bos}$, which depends on the bosonic polarization operator $\Pi$ (the bosonic
         self-energy)
         and is constructed such that the stationary condition
         $\delta F/\delta \Pi =0$  reproduces the
          conventional
          diagrammatic series for $\Pi$.

         We analyze
         the
         electron-phonon model and
         different
         electronic models
         in
         which a certain collective
         bosonic mode
         becomes massless at a QCP. For these models, the low-energy behavior of fermions and their soft collective excitations  is captured within an effective fermion-boson model, in which a collective mode becomes an independent degree of freedom, coupled to fermions.

         The full variational free energy of fermion-boson model is
          $F = F_{bos} + F_{el} + F_{int}$, where
$F_{int}$ is the sum of skeleton diagrams.
 We assume, following earlier works, that both phonons and soft collective modes are slow compared to dressed fermions,
  either because a velocity of a boson is small compared to that of a dressed fermion,
    or because collective bosons are Landau overdamped, and that the smallness of an (effective) velocity of a boson is controlled by a dimensionless parameter $\lambda_E$, often called Migdal-Eliashberg parameter (more on this below).
  In practical terms, the fact that
  the
  bosons are slow compared to fermions means that corrections to fermion-boson vertex
   are small as in the processes identified with vertex corrections fermions are forced to vibrate at boson frequencies,
    far away from their own resonance.  This makes higher-loop terms in the skeleton loop expansion of $F_{int}$ small compared to the one-loop term, and we keep only this term in $F_{int}$.

The  free energy
of the $\gamma-$model
 is
  $F_\gamma = F_{el} + F_{int}$.  Like we said, the specific heat obtained from $F_\gamma$
 depends on the cutoff.
  Our goal is to understand the role of $F_{bos}$, specifically
   (i) whether
  it  acts as the
  counter term,
  which eliminated the cutoff dependence of $F_{\gamma}$,
  and (ii)  whether it also affects the universal part of $C(T)$.

We show below  that for any $\gamma$, the full free energy $F$ near a QCP is
  \beq
  F =
  -2\pi T N_F \sum_m \rvert \omega_m \rvert
  + \frac{T}{2} \sum_q  \log [{-D^{-1}_q}]
  \label{Fb_a}
  \eeq
 where the first term is the contribution from free fermions, and in the second $q \equiv ({\bf q}, \Omega_m)$, $\Omega_m = 2\pi T m$, and
 $D_q$ is the fully dressed bosonic propagator.
 This result holds even if we include thermal fermionic self-energy, which near a QCP has to be computed self-consistently beyond Eliashberg theory~\cite{avi}.
 The fermion-boson interaction is present in
 $D_q$
  as it contains  the bosonic polarization bubble.
 We compute the specific heat from (\ref{Fb_a})
 and
 compare
 it
 with the one of the regularized $\gamma-$model.
 We then
 address the issues (i) and (ii).
  Regarding
  (i), we find  that $F_{bos}$ cancels the cutoff-dependent terms in $F_\gamma$, i.e.
  it provides
  the physical realization of the
  counter
  term.
On (ii), the result depends on whether $\gamma<1$ or $\gamma >1$.  For $\gamma <1$ (Ising-nematic and related models),
the contribution from $F_{bos}$ to the universal part of the specific heat is of the same order as the one from the regularized $F_\gamma$.
The two contributions
differ by a $\gamma-$dependent factor, e.g., by $3/2$ for $\gamma =1/3$.
For $\gamma >1$, including the electron-phonon case ($\gamma=2$), the contribution from $F_{bos}$ to the universal part of $C(T)$
coincides with that from free bosons (a $T$-independent term for $\gamma =2$).
The full $F$ in this case (Eq.  (\ref{Fb_a})) is the sum of contributions from free bosons
with the dressed mass
and from the regularized $\gamma$ model. The last contribution is negative at small $T$, in agreement with Refs. \cite{yuz_2,yuz_3}.  However, the negative term appears in Eq. (\ref{Fb_a}) as
the subleading term in the expansion
        $\log [{-D^{-1}_q} ] $ to first order in the dynamical part of the bosonic polarization, while
        the positive contribution from free bosons
         is the leading term. The expansion holds in powers of the Migdal-Eliashberg parameter  $\lambda_E$ (defined below), and we argue that as long as $\lambda_E \leq 1$, i.e., as long as the theory is under control, the
          specific heat is positive.
           Based on this, we argue that the normal state
           remains stable at a QCP and at any distance away from it.
            In this last respect our conclusions are different from those in Refs. \cite{yuz_2,yuz_3}.

 The structure of the paper is
 as follows.
  In Sec. \ref{sec:free} we present  the generic Luttinger-Ward-Eliashberg expression for the free energy, briefly discuss the Eliashberg theory, and use it to obtain the expressions for the full $F$,
   Eq.~(\ref{Fb_a}),
   and for $F_\gamma$ in the purely fermionic $\gamma-$model.
   In
   Sec.~\ref{sec:1/3}
   we compare the two expressions for the Ising-nematic model in 2D. Here we also show that the result for $F$ does not change if we include thermal self-energy, which has to be calculated outside the Eliashberg theory, and estimate the strength of vertex corrections once we include thermal self-energy.  In
   Sec.~\ref{sec:1/2}
   we consider antiferromagnetic QCP in 2D.  In
   Sec.~\ref{sec:2}
   we consider an electron-phonon system near a QCP.  Here we also discuss, in Sec.~\ref{sec:new},
  the regularization of $F_\gamma$ from physical perspective.  In Sec.~\ref{sec_ext} we extend the $\gamma =2$ model
   to arbitrary $\gamma$ between 1 and 2 and compute the specific heat. We show that the full specific heat is positive,  as long as $\lambda_E \leq 1$,  despite that the contribution from the regularized $F_\gamma$ is negative.  We present our conclusions in Sec. \ref{sec:concl}.   Some technical details of the calculations are presented in the Appendices.

 \section{Free energy and specific heat}
 \label{sec:free}

 The variational free energy  for interacting fermions
  has been derived by Luttinger and Ward~\cite{lw} and extended to fermion-boson systems by Eliashberg~\cite{Eliashberg} (see also~\cite{bs,pk}). For more recent studies of variational free energy,
  see Refs.
\cite{haslinger,cmg,*cmg_long,secchi2020phonon,benlagra2011luttinger,boyack,yuz_1,zhang_22}.
  The free energy per unit volume is the sum of
  the fermionic contribution, the bosonic contribution, and the contribution due to fermion-boson interaction:
 \beq
  F = F_{el} + F_{bos} + F_{int}.
 \label{F}
 \eeq
  The fermionic part is
  \beq
  F_{el} =
  -2 T \sum_k \log{G^{-1}_k}
   + 2 i T \sum_k \Sigma_k G_k
  \label{Fe}
  \eeq
  where $k \equiv ({\bf k}, \omega_m)$, $\omega_m = \pi T (2m+1)$, $T \sum_k  = T  \sum_m \int d{\bf k}/(2\pi)^d$ ($d$ is the spatial dimension),
   ${\tilde \Sigma}_k = \omega_m + \Sigma_k$, where $\Sigma_k$ is the self-energy,  and $G_k =  (
   i
   {\tilde \Sigma}_k -\epsilon_{\bf k})^{-1}$ is the Green's function.

The bosonic part is
 \beq
F_{bos} = \frac{T}{2} \sum_q \left( \log [{-D^{-1}_q}]  +  \Pi_q D_q\right)
  \label{Fb}
  \eeq
 where $q \equiv ({\bf q}, \Omega_m)$, $\Omega_m = 2\pi T m$, $\Pi_q$ is the bosonic self-energy, and
 $D_q = ((D^{0}_q)^{-1} - \Pi_q)^{-1}$
is the dressed bosonic propagator.
For the bare bosonic propagator we set
$D^0_q = -D_0/(\Omega^2_m + \omega^2_D)$ for the electron-phonon case,
where $\omega_D$ is a bare Debye frequency,
$D^0_q = -D_0 /((\Omega_{m}/c)^2 + {\bm q}^2+m^2)$  for the Ising-nematic case,
and
$D^0_q = -D_0/((\Omega_{m}/c)^2 +({\bm q}-{\bm Q})^2+m^2)$  for the antiferromagnetic case, where ${\bm Q} = (\pi,\pi)$,
$c$
is of order of Fermi velocity
$v_F$, and
$m$ is a bare boson mass.
We
set the lattice constant $a=1$.
For the last two cases,
the
$\Omega^2_m$ term in the bosonic propagator can
be neglected as
for relevant $\Omega_m$
it
is parametrically smaller than the Landau damping term from $\Pi_q$.
  On the contrary, for the electron-phonon case, $\Omega^2_m$ term is more relevant than the Landau damping.

 Finally,  the interaction part is
\beq
F_{int} = - T^2 \sum_{k,k'} g^2_{|{\bf k}-{\bf k}'|} G_k D(k-k') G_{k'} + ....
 \label{Fint}
\eeq
 where $g_{\bm q}$
 is the Yukawa coupling.
 The dots in (\ref{Fint})  stand for higher-order contributions, which account for vertex corrections
  (see Ref. \cite{cmg,*cmg_long,maslov4} for the discussion on higher-order terms in the loop expansion of $F_{int}$)
 We assume, following~\cite{Eliashberg}, that vertex corrections can be neglected (more on this below).
 For
 simplicity, we also
 approximate $g_{\bm q}$ by $g$.
 We
 refer to
 the free energy described by Eqs. (\ref{F}) - (\ref{Fint}) as
 the
 Eliashberg free energy.

 \begin{figure}
  \includegraphics[width=9cm]{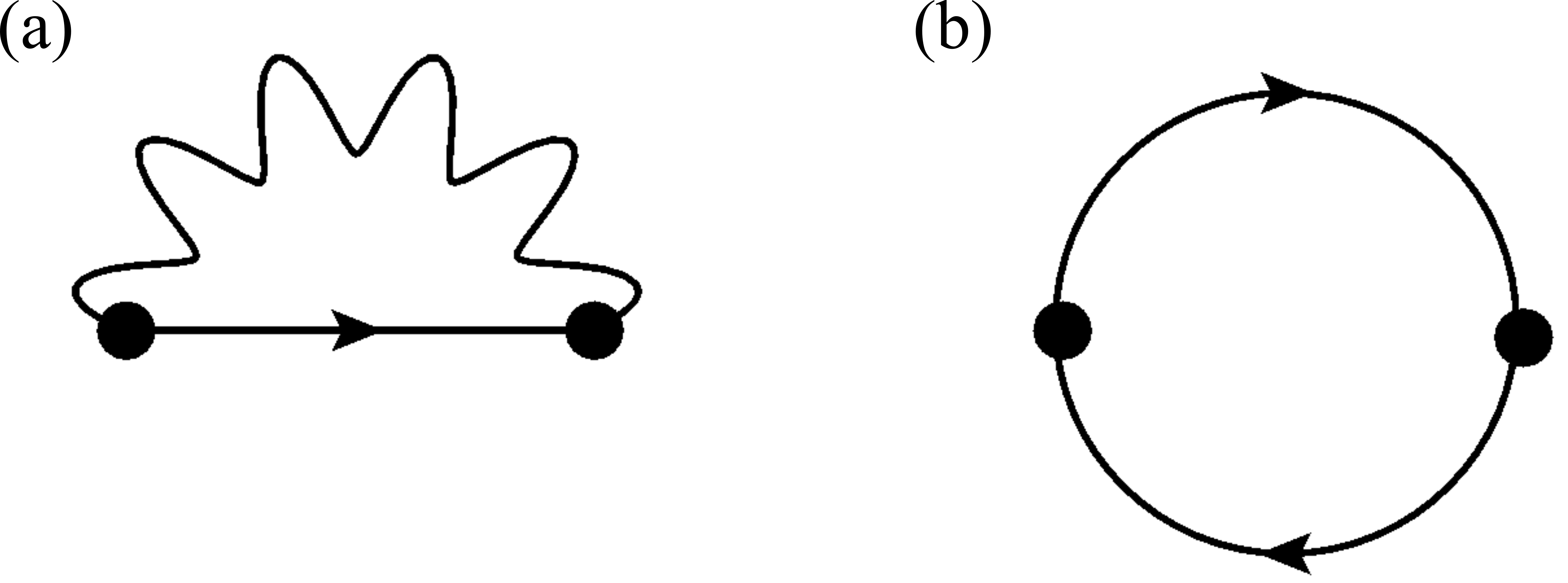}
  \caption{Self-energy of (a) electron $i\Sigma(k)$ and (b) boson $-\Pi(q)$ (the polarization bubble). The solid (waggle) lines denote the dressed electron (boson) Green's functions.
   The polarization bubble contains factor of $2$ from the spin degeneracy.}
  \label{fig:selfE}
\end{figure}

The stationary solutions for $\Sigma_k$ and $\Pi_k$ are obtained from $\delta F/\delta \Sigma_k =0$ and $\delta F/\delta \Pi_q =0$. They give rise to two Eliashberg equations for fermionic and bosonic self-energies~\cite{lw,Eliashberg,haslinger,cmg,*cmg_long,maslov4,boyack,yuz_1} (see Fig.~\ref{fig:selfE})
\bea
\Sigma_k &=& - i T \sum_{q} g^2 G_{k-q} D_{q}, \label{El_s} \\
\Pi_q &=& 2 g^2 T \sum_k G_k G_{k-q}, \label{El_p}
\eea
where the factor $2$ in Eq.~(\ref{El_p}) accounts for the spin degeneracy.
These equations are the same as one obtains diagrammatically, without invoking the free energy.
 We emphasize in this regard that the diagrammatic loop expansion  with full $G$ and full $D$ holds only for $F_{int}$.
  The terms $F_{el}$ and $F_{bos}$ are additional contributions to the free energy, constructed to
 reproduce Eqs. (\ref{El_s}) and (\ref{El_p}) as stationary conditions for the full $F$.

\begin{figure}
  \includegraphics[width=9cm]{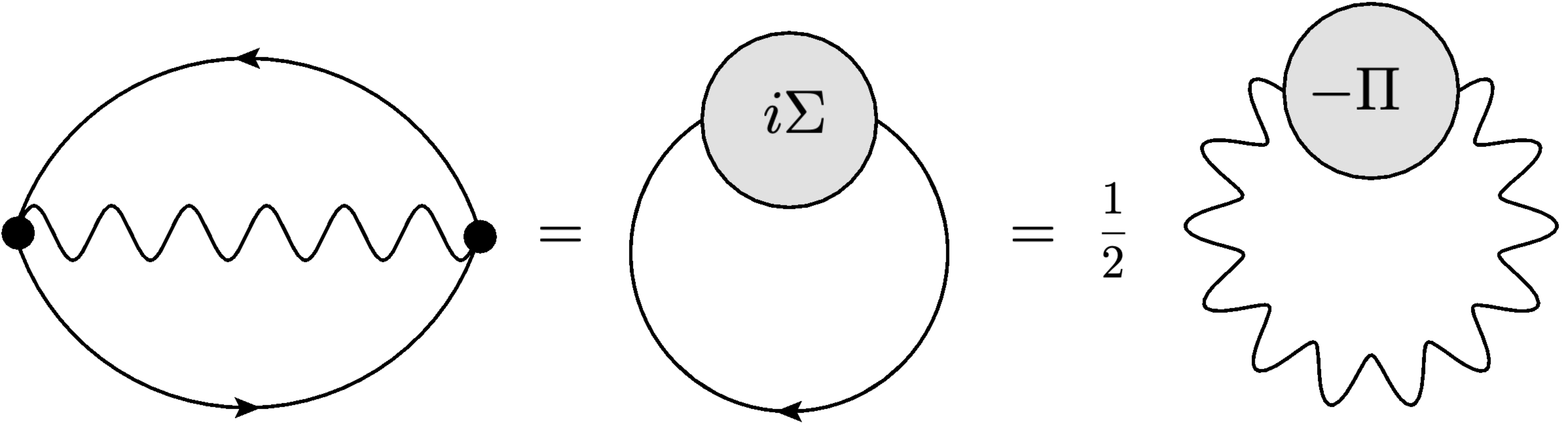}
  \caption{Thermodynamic potential due to fermion-boson interaction. The three diagrams are equivalent if we substitute the self-energy $\Sigma$ and the polarization $\Pi$ in Eqs. (\ref{El_s}) and (\ref{El_p}).}
  \label{fig:potential}
\end{figure}

Below we will analyze free energy in equilibrium, when $\Sigma_k$ and $\Pi_q$ obey Eqs. (\ref{El_s}) and (\ref{El_p}).  One can easily
check
that in this situation
\beq
 T/2 \sum_q \Pi_q D_q = i T \sum_k \Sigma_k G_k
 \label{eqn_5}
 \eeq
 because both expressions describe the same skeleton diagram, see
 Fig.~\ref{fig:potential}.  Along the same lines,
 \beq
F_{int} = - i T \sum_k \Sigma_k G_k.
\label{eqn_1}
\eeq
Using these two expressions, we obtain
  \beq
  F = -
  2T \sum_k \log{ G^{-1}_k}
   + 2 i T \sum_k \Sigma_k G_k
   +\frac{T}{2} \sum_q \log [-{D^{-1}_q}]
 \label{Fe_8}
  \eeq
and separately
  \beq
  F_{el} + F_{int} = -
  2T \sum_k \log{G^{-1}_k}
  +  i T \sum_k \Sigma_k G_k
  \label{Fee}
  \eeq

\subsection{Eliashberg theory}

The Eliashberg formula for the free energy is valid when bosons are slow
compared to
the
fermions, either because $\omega_D
\ll E_F$
in the electron-phonon problem,
or because
the
collective boson is Landau overdamped. An extension to $N \gg 1$ fermionic flavors, which individually interact with a boson, enhances the magnitude of
the
Landau damping
term
and
increases the applicability range of the Eliashberg theory
~\cite{Altshuler1994,millis_92,acs,raghu_15,*Wang_H_17,*Wang_H_18,*Fitzpatrick_15,rech_2006,*rech_2006_1,Chowdhury_2020,esterlis_21,guo_22}.

\begin{figure}
  \includegraphics[width=7cm]{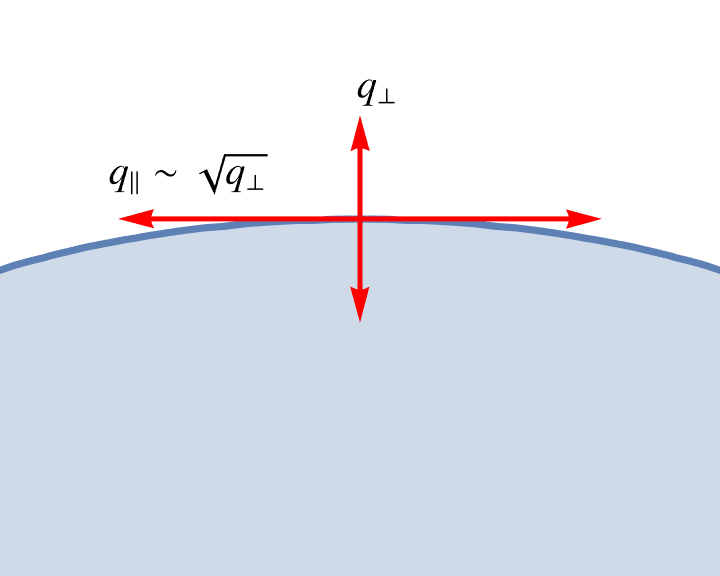}
  \caption{Typical transverse  and longitudinal momenta, $q_{\bot}$ and $q_{\parallel}$, for the Ising-nematic case. At small ${\bf q}$,  $q_{\parallel} \gg q_{\bot}$.}
  \label{fig:FS_IN}
\end{figure}

 The condition that
 the
 bosons are slow compared to
 the fermions
 allows one to factorize the
  momentum
  integration along and transverse to the Fermi surface because in all three cases that we consider,
  the
  typical transverse momenta  are much smaller than typical longitudinal momenta
(we illustrate this in  Fig.~\ref{fig:FS_IN}).
  To obtain the leading contribution to the r.h.s. of (\ref{El_s}) one can then integrate over transverse $q_{\perp}$ in the fermionic propagator  and over $q_{\parallel}$ in the bosonic propagator with ${\bf q}$ connecting points on the Fermi surface~\cite{haslinger}.
Integrating over momenta this way
 and  extending both integrations to infinity
\footnote{We assume, as in previous works on metallic QCP, that fermionic bandwidth $W$ is the largest scale of the problem,
 and neglect terms, which are  small in $g/W$.},
  one obtains a purely dynamical self-energy
\beq
\Sigma_k = \Sigma (\omega_n) =  \pi T \sum_m {\text{sgn}} (\omega_n + \Omega_m) D_{loc} (\Omega_m),
\label{sigma_el}
\eeq
At $T=0$,
\beq
 \Sigma (\omega_n) =  \int_0^{\omega_n} D_{loc} (\Omega_m)  d\Omega_m
\label{sigma_el0}
\eeq
 The form of $D_{loc}$ is model-specific, but in all cases we have  at a QCP
\beq
D_{loc} (\Omega_m) = \left(\frac{\bar g}{|\Omega_m|} \right)^{\gamma}
\label{dloc}
\eeq
 where $\gamma =1/3$ for the Ising-nematic case, $\gamma =1/2$ for the antiferromagnetic case, and $\gamma =2$ for the electron-phonon case.
 The coupling ${\bar g}$ is expressed via $g$ (
 see Secs.
 \ref{sec:1/3},\ref{sec:1/2},\ref{sec:2} below).
Away from
the
QCP, Eq. (\ref{dloc}) is modified to
  \beq
D_{loc} (\Omega_m) = \left(\frac{\bar{g}^2} {\Omega^2_m + M^2} \right)^{\gamma/2}
\label{dloc_1}
\eeq
where $M \sim m^3$ for Ising-nematic case, $M\sim m^2$ for antiferromagnetic case, and $M = \bar{\omega}_D$ (renormalized Debye frequency) for the electron-phonon case.
 We assume that $M$ and $\omega_D$ do not depend on temperature, or, more accurately, that their temperature dependence yields smaller $C(T)$ compared to what we find below.

Because $\Sigma_k$
 in (\ref{sigma_el}) does not depend on
 $\epsilon_{\bm k}$,
 one can explicitly integrate over
 momentum in Eq.  (\ref{Fe}) using $\int d^dk/(2\pi)^d = N_F \int d
 \epsilon_{\bm k}$,
  where $N_F$ is the density of states at the Fermi level
per spin component.
The integration yields
\bea
&&
-2T \sum_k \log{G^{-1}_k}
 = -2\pi T N_F \sum_m \left( |\omega_m| + |\Sigma (\omega_m)|\right) \nonumber \\
&&2 i T \sum_k \Sigma_k G_k = 2\pi T N_F \sum_m |\Sigma (\omega_m)|
\label{eqn_3}
\eea
Combing the two contributions, we find that the self-energy cancels out and $F_{el}$ retains  the same as for free fermions:
\beq
F_{el} = -2\pi T N_F \sum_m |\omega_m|
\label{eqn_2}
\eeq
We emphasize that this holds only if $\Sigma_k$ does not depend on
$\epsilon_{\bm k}$.
For a generic momentum and frequency dependent $\Sigma_k$,  $F_{el}$ does depend on the fermionic self-energy.

Applying the same procedure to Eqs. (\ref{Fe_8}) and (\ref{Fee}) we obtain
~\cite{haslinger,cmg,*cmg_long,yuz_1,zhang_22}
\beq
  F = -2\pi T \sum_m |\omega_m|
   +\frac{T}{2} \sum_q \log[-{D^{-1}_q}] = F_{free} +\frac{T}{2} \sum_q \log[-{D^{-1}_q} ]
 \label{Fe_9}
  \eeq
  and
 \beq
  F_{el} + F_{int} = -2\pi T N_F \sum_{m} |\omega_m|  - \pi T N_F \sum_{m} |\Sigma (\omega_m)|
  \label{Fe_1}
  \eeq
 Note that the self-energy $\Sigma (\omega_m)$ cancels out in $F$, and that the dependence on fermion-boson interaction comes about because  $D_q$ depends on the polarization $\Pi (q)$.

 At $T=0$,  $\Sigma (\omega_m) = ({\bar g}^{\gamma}/(1-\gamma)) |\omega_m|^{1-\gamma} {\text{sgn}} \omega_m
 \equiv
 \omega_0^{\gamma}|\omega_m|^{1-\gamma} {\text{sgn}} \omega_m$, where
$\omega_0 ={\bar g}/(1-\gamma)^{1/\gamma}$. This holds for $\gamma <1$. For $\gamma >1$, one has to add
 the contribution from the lower limit in (\ref{sigma_el0}). This last contribution scales as $1/M^{\gamma -1}$ and diverges at $M \to 0$. However, it does not contribute to the specific heat, as one can explicitly verify.  At a finite $T$, the self-energy becomes a function of a Matsubara number, and there appears a
 separate singular contribution $O(1/M^{\gamma})$ from zero bosonic Matsubara frequency.  This last contribution requires special attention, and we discuss it in some detail in Sec. \ref{sec:1/3}.

\subsection{A purely electronic $\gamma$-model}

The $\gamma$-model is
designated to reproduce
some low-energy properties of
the
fermion-boson system
(more specifically,
non-FL
and superconductivity).
It is a
fermion-only model in which $D_{loc} (\Omega_m)$ from
 (\ref{dloc_1})
plays the role of an  effective dynamical 4-fermion
 interaction~\cite{paper_1,*paper_2,*paper_3,*paper_4,*paper_5,*paper_6,*paper_odd}.
 The model allows one to
analyze the interplay between non-FL and pairing by solving coupled Eliashberg  equations for the dynamic fermionic self-energy and the dynamic pairing vertex $\Phi (\omega_m)$.
In
a more common and
convenient  formulation, these equations are re-expressed in terms of the superconducting gap function $\Delta (\omega_m)$ and the
inverse
quasiparticle residue $Z (\omega_m)$.  By construction, the model contains only the fermions,  and its free energy in the normal state is $F_\gamma =  F_{el} + F_{int}$, given by (\ref{Fe_1}):
  \beq
  F_\gamma = -2\pi T N_F \sum_{m} |\omega_m|  - \pi^2 T^2 N_F {\bar g}^\gamma \sum_{m,m'}
   \frac{{\text{sgn}} (\omega_m \omega_{m'})}{((\omega_m - \omega_{m'})^2 +
   M^2)^{\gamma/2}}
  \label{Fe_2}
  \eeq
   The  summation over $m$ is confined to frequencies below the upper energy cutoff $\Lambda$ of the $\gamma-$model.
   In practice, this implies that the summation holds over $M_f$ positive and $M_f$ negative fermionic Matsubara frequencies $\omega_n = \pi T (2n+1)$  ($-M_f <n < M_f-1)$. The relation between $M_f$
    and  $\Lambda$ can be obtained by comparing the exact sum
     of $|\omega_m|$ with Euler-Maclauren formula, in which the integral is cut by $\Lambda$. The comparison yields
    ~\cite{zhang_22}
    $4 \pi^2 T^2 M^2_f = \Lambda^2 + \pi^2 T^2/3$, hence
  \beq
   M_f = {\tilde \Lambda} \left(1 + \frac{1}{24 {\tilde \Lambda}^2 }+ ..\right)
   \label{Fe_ex_1}
   \eeq
    where ${\tilde \Lambda} = \Lambda/(2\pi T)$.

   Applying this procedure to both terms in (\ref{Fe_2}), we obtain~~\cite{zhang_22} at $T \gg M$
   \bea
  F_\gamma &=& -N_F \left(\Lambda^2 - {\bar g}^\gamma \Lambda^{2-\gamma} \frac{2(1-2^{-\gamma})}{(1-\gamma)(2-\gamma)}\right) \nonumber \\
  &-& N_F \pi T \Lambda \left(\frac{\bar g}{
  M}\right)^\gamma -  N_F  \Lambda {\bar g}^\gamma (2\pi T)^{1-\gamma} \zeta (\gamma) \nonumber \\
  & +& N_F \left(\frac{3}{2} {\bar g}^\gamma (2\pi T)^{2-\gamma} \zeta (\gamma-1) - \frac{1}{3} \pi^2 T^2\right),
  \label{Fe_3}
  \eea
  where $\zeta(s)$ is the Riemann zeta function.
  The first two terms in (\ref{Fe_3}) constitute the free energy at $T=0$. The next  one, with
   $M$ in the denominator, comes from
   the
   thermal piece in $\Sigma (\omega_m)$ in (\ref{Fe_1}), or, equivalently, from
   the
   $m=m'$ term in (\ref{Fe_2}). The next term
   comes from $\omega_m, \omega_{m'} \sim \Lambda$, but $\omega_m - \omega_{m'} = O(T)$. The last term is the combination of cutoff independent contributions from both terms in (\ref{Fe_2}).

   The specific heat $C_\gamma (T) = -T d^2F_\gamma/d^2T$ is
   \bea
  C_\gamma (T) &=& 2\pi \Lambda N_F \left(\frac{\bar g}{2\pi T}\right)^\gamma \gamma (\gamma -1) \zeta (\gamma) \nonumber \\
  &&+ \frac{2}{3} \pi N_F \left(\pi T - \frac{9}{4} {\bar g}^\gamma (2\pi T)^{1-\gamma} (\gamma-2) (\gamma-1) \zeta (\gamma-1) \right)
    \label{Fe_4}
  \eea
 The first term in
 (\ref{Fe_4})
 is parametrically larger than the other two
 since it is proportional to $\Lambda$.
  This term is positive, but depends linearly on the upper energy cutoff.  The second term is a universal  contribution to $C(T)$.
   This term is positive for $\gamma <1$, but  becomes negative at small $T <T_0 = [{\bar g}/ (2\pi)] [9(\gamma-2)(\gamma-1)\zeta(\gamma-1)/2]^{1/\gamma}$ for $\gamma >1$, when  $(\gamma-2) (\gamma-1) \zeta (\gamma-1) >0$.
The temperature $T_0$ increases with $\gamma$ up to $\gamma =3$.

The authors of \cite{yuz_1,yuz_2,yuz_3} argued that the dependence of $C(T)$ on the cutoff is
spurious
and must be eliminated by a proper regularization. They suggested that this
is achieved by  adding to the r.h.s. of (\ref{Fe_2}) the term
\beq
\pi^2 T^2 N_F {\bar g}^\gamma \sum_{m,m'}
   \frac{1}{((\omega_m - \omega_{m'})^2 +
   M^2)^{\gamma/2}}
 \label{Fe_5}
 \eeq
 This additional term cancels out all $\Lambda$-dependent terms in $F_\gamma$ in (\ref{Fe_3}) and changes the prefactor for the universal $T^{2-\gamma}$ term.  The
  regularized free energy, which we label as ${\bar F}_\gamma$, is
  \bea
  {\bar F}_\gamma = N_F \left({\bar g}^\gamma (2\pi T)^{2-\gamma} \zeta (\gamma-1) - \frac{1}{3} \pi^2 T^2\right)
  \label{Fe_6}
  \eea
 This yields a universal, cutoff-independent specific heat ${\bar C}_\gamma (T)$.   At a QCP,
  \beq
  {\bar C}_\gamma (T) =  2\pi N_F \left(
  \frac{\pi}{3} T -  {\bar g}^\gamma (2\pi T)^{1-\gamma} (\gamma-2) (\gamma-1) \zeta (\gamma-1)\right)
  \label{Fe_7}
  \eeq
For $\gamma <1$, all terms in (\ref{Fe_7}) are positive as
$\zeta (\gamma-1)$
is negative.
For
$\gamma =1/3$,
\beq
  {\bar C}_{1/3} (T) =  2\pi N_F \left(
  \frac{\pi}{3} T  + 0.172 {\bar g}^{1/3} (2\pi T)^{2/3}\right)
  \label{Fe_7_1}
  \eeq

  For $\gamma >1$,  $C(T)$ given by (\ref{Fe_7}) is still negative at small $T$ because $(\gamma-2) (\gamma-1) \zeta (\gamma-1) >0$.   For $\gamma \to 2$,
  \beq
  {\bar C}_2 (T) =  2\pi N_F \left(
  \frac{\pi}{3} T -  \frac{{\bar g}^2}{2\pi T}\right)
  \label{Fe_7_2}
  \eeq
  
  \subsection{Underlying fermion-boson model}

We now return back to the underlying fermion-boson model, in which there is an additional bosonic contribution to the free energy, and
check
whether the effect of $F_{bos}$  is the same as of the extra term (\ref{Fe_5}), which regularizes $F_\gamma$.

The full free energy $F = F_{el} + F_{bos} + F_{int} = F_\gamma + F_{bos}$
is given by Eq. (\ref{Fe_9}) as the sum of
the
free-fermion contribution and the one expressed via the full bosonic propagator.
In contrast,
$F_\gamma$, given by Eq. (\ref{Fe_1}), depends
explicitly
on the fermionic self-energy.
  We now
  study the relation between these two expressions.
  We show that the outcome depends on the type of a QCP. To see this, we consider separately Ising-nematic QCP,  antiferromagnetic QCP, and a QCP of an electron-phonon system.

\section{Ising-nematic QCP}
\label{sec:1/3}

We consider a 2D
system.
The bare bosonic propagator has the
 Ornstein-
 Zernike
 form
$D^{0}_q = - D_0/(q^2 + m^2)$.
The static part of  $\Pi (q)$ renormalizes $D_0$ and $m$.
We assume that these renormalizations are already incorporated into $D^{0}_q$.
The dynamical part of  $\Pi (q)$ accounts for the Landau damping: $\Pi ({\bf q}, \Omega_m) - \Pi ({\bf q},0) =  (1/D_0) \alpha  |\Omega_m|/|{\bf q}|$. For a circular Fermi surface, $\alpha =  g^* k_F/(\pi v^2_F)$, where
$g^* = g^2 D_0$
has the dimension of energy and plays the role of an effective fermion-boson interaction.
 The dressed bosonic propagator is
\beq
 D_q = - \frac{D_0}{ \rvert {\bm q} \rvert^2 + m^2+ \alpha \frac{|\Omega_m|}{\rvert {\bm q} \rvert}}.
 \label{Fe_10}
 \eeq
 Integrating over one momentum component and comparing with $D_{{\text{loc}}} (\Omega_m)$ from (\ref{dloc}) for $\gamma =1/3$, we obtain ${\bar g} = (g^*)^2/(162 \sqrt{3} \pi^2 E_F)$ and $\omega_0 = (27/8) {\bar g} =(g^*)^2/(48 \sqrt{3} \pi^2 E_F)$. The mass $m$ is related to $M$ in the $\gamma-$model by $M = 32\pi/(81 \sqrt{3}) (mv_F)^3/(g^* E_F)$.

Substituting $D_q$  into (\ref{Fe_9}), subtracting from $\log [-{D^{-1}_q}]$
its static part, which does not contribute to the specific heat, and integrating over momentum (the integral converges),  we obtain at a QCP (i.e., at $m=0$)
 \bea
 F = F_{I-N} &=& F_{\text{free}} + \frac{\alpha^{2/3}}{4\pi \sqrt{3}}  (2\pi T)^{5/3} \sum_{1}^{M_b}  n^{2/3}  \nonumber  \\
 &=& F_{\text{free}} + \frac{\alpha^{2/3}}{4\pi \sqrt{3}}  (2\pi T)^{5/3} H_{-2/3}  (M_b)
 \label{Fe_11}
 \eea
 where
 $F_{\text{free}} =-N_F (\Lambda^2 + \pi^2 T^2/3)$ is free energy of a gas of free fermions, and
 $H_p (M_b) =\sum_{1}^{M_b} 1/k^p$ is the Harmonic number. The asymptotic expansion of $H_p (M_b)$ at large $M_b$  is
 \beq
H_p (M_b) = \frac{\left(M_b+\frac{1}{2}\right)^{1-p}}{1-p} + \zeta (p) + O(1/M_b^{p+1})
\eeq
The relation between $M_b$ and and the cutoff $\Lambda$ can be established in a way similar to
the procedure described above
for fermions, by evaluating $\sum_{m=1}^{M_b} m$ directly and using Eular-Maclaurin formula with $\Lambda$ as the upper cutoff of frequency integration.  This yields
\beq
M_b + \frac{1}{2} = {\tilde \Lambda} \left(1 + \frac{1}{24 {\tilde \Lambda}^2 }+ ..\right).
\label{Fe_ex_2}
\eeq
Substituting into (\ref{Fe_11}), we obtain
\bea
 F_{I-N} &=& -N_F \Lambda^2 + \frac{\sqrt{3}(\alpha A)^{2/3}}{20\pi} \Lambda^{5/3} \nonumber \\
 && -  \frac{\pi^2}{3} N_F T^2 + \frac{\alpha^{2/3}}{4\pi \sqrt{3}} (2\pi T)^{5/3} \zeta (-2/3)
\label{Fe_12}
\eea
where $\zeta (-2/3) \simeq -0.155$.
Differentiating
with respect to
$T$, we find that both
the
entropy $S_{I-N}(T) = - dF_{I-N}/dT$ and
the
specific heat $C_{I-N}(T) =- T d^2F_{I-N}/dT^2 =(2/3) S_{I-N}(T)$ are
independent on $\Lambda$.
  The specific heat is
\beq
C_{I-N}(T) = \frac{2\pi^2}{3} N_F T - \frac{5 \alpha^{2/3}}{9 \sqrt{3}}(2\pi T)^{2/3} \zeta (-2/3)
\label{Fe_14}
\eeq
Re-expressing the result in terms of ${\bar g}$ from Eq. (\ref{dloc}),
we obtain
\beq
C_{I-N}(T) = \frac{2\pi^2}{3} N_F T - \frac{5}{3} {\bar g}^{1/3} (2\pi T)^{2/3} \zeta (-2/3)
\label{Fe_14_1}
\eeq
Comparing this $C_{I-N} (T)$ with the ${\tilde C}_{1/3} (T)$ from (\ref{Fe_7_1}) (a regularized specific heat in the $\gamma-$model), we see that they agree up to a numeric prefactor in the $T^{2/3}$ term (the one in $C_{I-N} (T)$ is larger by $3/2$.
 The factor $3/2$ is the difference between the momentum integral of $\log(-{D^{-1}_q})$ with static term subtracted and
  of $\Pi_q D_q$, i.e., between $\int_0^\infty dx x \log({1+1/x^3}) =\pi/\sqrt{3}$ and $\int_0^\infty dx/(x^3+1) =(2/3) (\pi/\sqrt{3})$).

The analysis at Ising-nematic QCP can be formally extended to other
values of
$\gamma$ if we replace $q^2$ in $D_q$ by $q^a$
 with some $a >1$.  The exponent $\gamma$ then changes from $1/3$ to $\gamma = (a-1)/(a+1)$, which ranges between $0$ and $1$.
 One can easily verify (see Appendix~\ref{sec:extension_app})
 that the interaction contributions to $C _{I-N}(T)$ in the Ising-nematic model and in the regularized $\gamma$-model have the same structure and just differ by $1-\gamma$ (the prefactor is larger in $C_{I-N} (T)$). The conclusion here is that for the Ising-nematic case the result of keeping the bosonic contribution to the specific heat is almost entirely reproduced by  either regularizing  the fermionic part of the
   free energy, as it was done in ~\cite{yuz_1,yuz_2,yuz_3}, or just eliminating the cutoff-dependent term in the specific heat in (\ref{Fe_4}).

  {\subsection{Role of thermal fluctuations}}

Eq.~(\ref{Fe_9}) for the free energy is obtained under the assumption that the momentum dependence of the self-energy  can be neglected for
$\epsilon_{\bm k} \sim {\tilde \Sigma} (\omega)$.
As mentioned above,
this is the case when
the
typical momenta transverse to the Fermi surface
 in Eq. (\ref{El_s})
are much smaller than typical momenta along the Fermi surface for the same frequency.
At $T=0$, this holds both at
the
QCP and away from it.  At
the QCP we have  $q^{\text{typ}}_{\perp} \sim {\tilde \Sigma} (\omega)/v_F$ and
$q^{\text{typ}}_{\parallel} \sim (\alpha \omega)^{1/3}$.
We use
${\tilde \Sigma} (\omega) = \omega + \omega^{1/3}_0 \omega^{2/3}$, where
$\omega_0  \simeq (g^*)^2/E_F \sim \bar{g}$.
A simple analysis shows that $q^{\text{typ}}_{\parallel} \gg q^{\text{typ}}_{\perp}$ up to
$\omega_{\text{max}} \sim (g^* E_F)^{1/2} \sim {\bar g}^{1/4} E^{3/4}_F$.
This scale is much larger than the upper cutoff for non-FL behavior,
$\omega_0 \sim {\bar g}$, which is also a typical scale for superconductivity.
Away from a QCP, typical $q_{\parallel}^{\text{typ}} \sim \text{max} \{m,(\alpha \omega)^{1/3} \}$ are even larger.

At a finite $T$ the self-energy $\Sigma_k$ can be split into two parts~\cite{acs,DellAnna2006,*metzner_new,punk,torroba_1,*torroba_2,avi}.
One is
the thermal part,
$\Sigma^{th}_k$, which comes from zero bosonic Matsubara frequency,
and the other is the quantum part,
$\Sigma^{q}_k$, which comes from all non-zero Matsubara frequencies.
For the quantum part, the condition $q^{\text{typ}}_{\parallel} \gg q^{\text{typ}}_{\perp}$
holds
up to
$\omega_{\text{max}}$, and
  one can evaluate $\Sigma^{q}$ using Eq.~(\ref{sigma_el}). For $T \gg M \sim m^3/\alpha$,
\beq
\Sigma^{q}_k = \text{sgn}(\omega_m) \left[ {3\over 2} {\bar g}^{1/3} \rvert \omega_m \rvert^{2/3} (1+O(T/\omega_m)) + \zeta(1/3) {\bar g}^{1/3} (2\pi T)^{2/3} \right].
\label{aa_1}
\eeq
 (see Ref. \cite{avi} for the analysis of $\Sigma^{q}_k$ at all $T$).

For the thermal part, the situation is different: the condition
$q^{\text{typ}}_{\parallel} \gg q^{\text{typ}}_{\perp}$ holds only away from a QCP, at a finite bosonic mass
$m > {\tilde \Sigma}_k/v_F$, where ${\tilde \Sigma}_k = \omega + \Sigma^{th}_k + \Sigma^{q}_k$.
Under this condition, we obtain
\beq
\Sigma^{th}_k = \text{sgn}(\omega_m) {g^* T \over 4 m v_F} = \text{sgn}(\omega_m) \pi T\left( { {\bar g}\over M} \right)^{1/3}.
\label{aa}
\eeq
A straightforward analysis shows that Eq. (\ref{aa}) is valid for
$ T < T^* \sim  (m/k_F)^2 E^2_F/g^*$.
At
the
QCP, $T^*$ vanishes, and at any finite $T$,
$q^{\text{typ}}_{\parallel} \ll q^{\text{typ}}_{\perp}$.
The thermal contribution to the self-energy then has to be computed differently, by integrating over both components of momenta in the fermionic propagator.  For the one-loop self-energy this yields
   \beq
   \Sigma^{th}_k = i B G_k = \frac{B}{{\tilde \Sigma}_k +i \epsilon_k}
   \label{Fe_15}
   \eeq
   where $B = g^* T \log ({k_F/m})$
   and ${\tilde \Sigma}_k = \omega + \Sigma^q_k + \Sigma^{th}_k = {\tilde \Sigma}^{q}_k + \Sigma^{th}_k$.
   The thermal self-energy diverges, but only logarithmically.  It has been argued~\cite{blaizot,torroba_1,*torroba_2,guo_22}
    that  the renormalization of  $m$ at a finite $T$ by high-energy fermions makes it $T$-dependent, in which case
    $m$ under the logarithm is cut by $T/v_F$, i.e., $\log ( {k_F/m} ) $ can be approximated by
    $\log ( {E_F/T} )$.
    We follow these works and set
    $B = g^* T \log ({E_F/T})$.

  \begin{figure}
  \includegraphics[width=16cm]{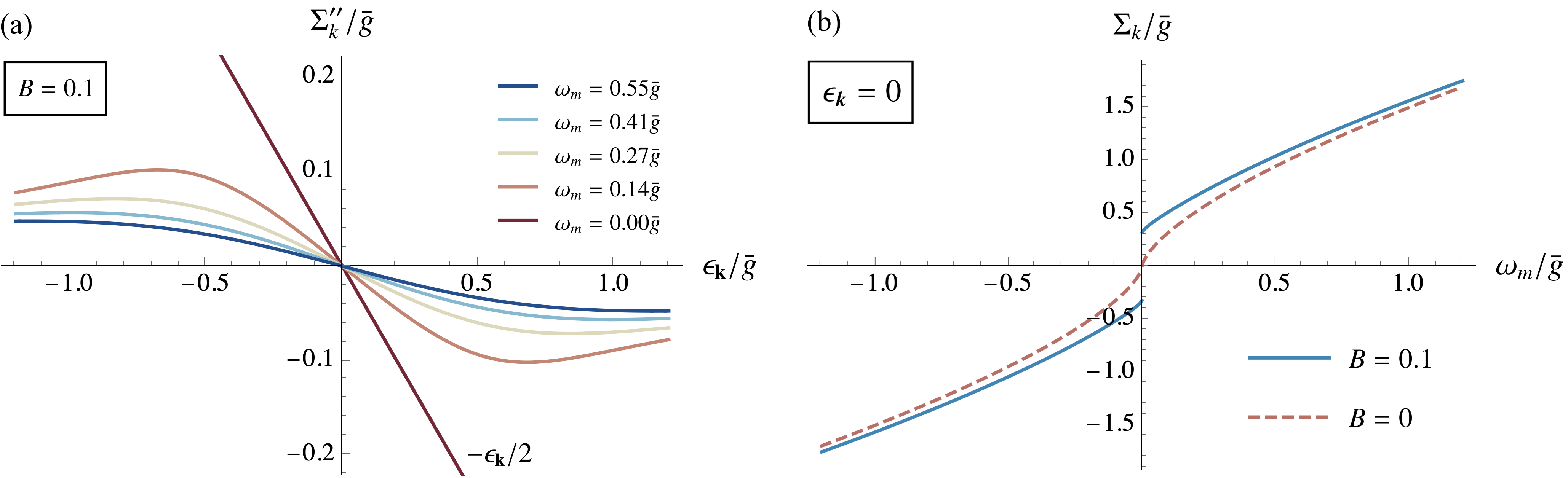}
  \caption{(a) The imaginary part of $\Sigma_k$ for different values of $\omega_m$ at $B=0.1$. (b)
   $\Sigma_k \equiv \Sigma'_k$ at $\epsilon_{\bm k} = 0$ at $B=0$ and  $B=0.1$ (dashed and solid lines, respectively).}
  \label{fig:self2}
\end{figure}

The key new feature of $\Sigma^{th}_k$ in (\ref{Fe_15}) is that it now depends on both $\omega_m$ and $\epsilon_{\bm k}$.  Then one has to redo the integration over $\epsilon_{\bm k}$ in the fermionic part of the free energy in Eq. (\ref{Fe_8}).  To do this, we solve Eq. (\ref{Fe_15}) for $\Sigma^{th}$ in terms of ${\tilde \Sigma}^q = \omega_m + { \Sigma}^{q} (\omega_m)$ and $\epsilon_{\bm k}$. We obtain
\beq
\Sigma^{th}_k = \sqrt{B + \left(\frac{{\tilde \Sigma}^q_k + i \epsilon_{\bm k}}{2}\right)^2} - \frac{{\tilde \Sigma}^q_k + i \epsilon_{\bm k}}{2}
\label{Fe_16}
\eeq
where we choose the branch cut of the square root along the negative real axis.
A similar expression, but at $\epsilon_{\bm k} =0$ and at ${\tilde \Sigma}^q_k \approx \omega_m$ has been obtained in~\cite{avi}.
In Fig.~\ref{fig:self2} (a), we plot the imaginary part of the total self-energy $\Sigma^{th}_k+\Sigma^q_k$ from Eqs.~(\ref{aa_1}) and (\ref{Fe_16})
as a function of $\epsilon_{\bm k}$
  for
  different $\omega_m$. The dependence is linear in $\epsilon_{\bm k}$ at
   the smallest $\omega_m$, with the universal slope $-1/2$.
    This renormalizes the dispersion to $\epsilon_{\bm k}/2$. At larger $\omega_m$,
     the renormalization of $\epsilon_{\bm k}$ becomes negligible. The crossover between the two regimes is at  $\omega_m \sim B^{3/4}$ at the smallest $T$, and at $\omega_m \sim B$ at  $T > (g^*)^3/E^2_F$.
 In Fig.~\ref{fig:self2} (b) we plot $\Sigma_k$
 at $\epsilon_{\bm k }=0$, where it is
 necessarily
  real.
 At the smallest $\omega_m \sim T$, $\Sigma_k$ scales as $\omega_m^{2/3}$ at $B=0$ and tends to a finite value  $\Sigma_k \approx \sqrt{B}$ at a finite $B$.

We now substitute $\Sigma^{th} (\omega_m,
\epsilon_{\bm k})$
from Eq.~(\ref{Fe_16}) into ${\tilde \Sigma}_k = {\tilde \Sigma}^q_k + \Sigma^{th}_k$ and then into Eq.~(\ref{Fe_8}) and explicitly integrate over $\epsilon_{\bm k}$.
After some algebra (see Appendix~\ref{sec:free_app} for details),
we obtain
that $\Sigma^{th}$ {\it cancels out} from both terms in Eq.~(\ref{Fe_8}) which contain fermionic self-energy. Namely,
\bea
&& -T \sum_k \log{[\epsilon^2_{\bf k} + {\tilde \Sigma}^2_k]} = -2\pi T N_F \sum_{n} |\omega_m + \Sigma^q (\omega_m)|
\nonumber \\
&& 2 i T \sum_k \Sigma_k G_k = 2\pi T N_F \sum_{n} |\Sigma^q (\omega_m)|
\label{Fe_17}
\eea

As a result
Eq. (\ref{Fe_9}) holds despite
that the thermal self-energy depends on $\epsilon_{\bm k}$.
For completeness, we verified explicitly that
the
momentum dependence of $\Sigma^{th}$ does not generate a
 significant
momentum dependence of $\Sigma^q$ (which still contains the full self-energy in the Green's function in the r.h.s.
 of (\ref{El_s})).

\subsection{Strength of vertex corrections}

The free energy in (\ref{Fe_8}) is obtained within
a
 self-consistent one-loop approximation, which neglects the vertex corrections.
Several authors argued~\cite{Altshuler1994,rech_2006,*rech_2006_1,sslee,metlitski2010quantum1,metzner_1,eberlein_16,pimenov_21} that at $T=0$ lowest-order vertex corrections are  generally of order one
(or of order $1/N$ in large $N$ theories), but
higher-order corrections are $O(1)$ even at large $N$ (Ref. \cite{sslee}) and furthermore are logarithmically singular~\cite{metlitski2010quantum1}, except for special cases~\cite{damia_19}.
The logarithms, however, likely modify the quasiparticle residue but not the exponent $\gamma =1/3$, and hence do not affect the  $T^{2/3}$ behavior.

In this section, we estimate the strength of vertex corrections at a finite $T$.
We recall that at a finite bosonic mass $m$, there is a range $T < T^*$,
where $q^{\text{typ}}_{\parallel} \gg q^{\text{typ}}_{\perp}$ and
the
thermal self-energy $\Sigma^{th} \sim g^* T/(m v_F)$ is obtained by factorizing the momentum integration between fermionic and bosonic propagators, and a range $T > T^*$,
where
the
factorization does not hold. In this
last
regime,
$\Sigma^{th}$ generally depends on
 ${\tilde \Sigma}^q$ and $\epsilon_{\bm k}$,
 and is of order $[g^* T \log ({E_F/T) }]^{1/2}$ when it is larger than
 ${\text {max}} \{ {\tilde \Sigma}^q, |\epsilon_{\bm k}| \}$. At $m =0$,
 $T^* =0$, and the last regime holds for all $T$.

 For $T < T^*$, a simple analysis shows that the leading
 vertex
 correction to fermion-boson coupling
$\delta g^*/g^* \sim T/T^*$, i.e., it remains small in the same $T$ range where one can factorize the momentum integration.
For $T >T^*$, a similar analysis shows that $\delta g^*/g^* \sim g^* T \log ({E_F/T}) / ({\tilde \Sigma_k})^2$. This vertex correction is at most of order one.
 It is then reasonable to expect that  thermal vertex corrections do not modify $C(T) \propto T^{2/3}$ at a QCP, and, moreover, the prefactor differs from the one in Eq. (\ref{Fe_14}) at most by a factor $O(1)$.

\section{Antiferromagnetic QCP in 2D}
\label{sec:1/2}

  \begin{figure}
  \includegraphics[width=8cm]{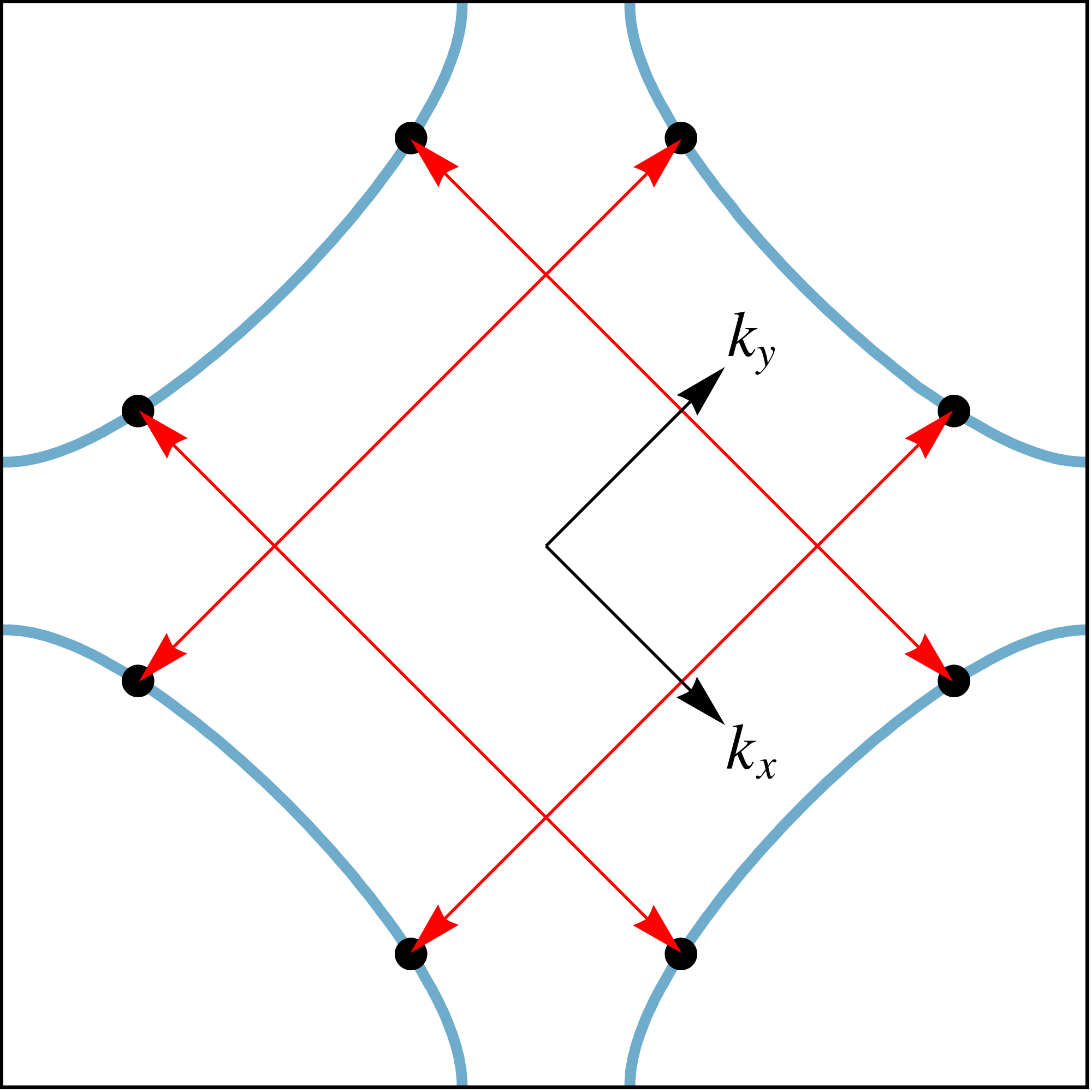}
  \caption{Fermi surface (blue lines) and hot spots (black dots), connected by the ordering wave vector ${\bf Q}$ (red arrows).  The coordinate frame $(k_x,k_y)$ is used in the main text.}
  \label{fig:fs_afm}
\end{figure}

The non-FL physics at a QCP towards spin order with momentum ${\bf Q} = (\pi,\pi)$  is described by the $\gamma$-model with $\gamma =1/2$,
 as the self-energy in hot regions  on the Fermi surface (the ones in which both
 $\epsilon_{\bm k}$ and $\epsilon_{{\bm k}+{\bm Q}}$ 
 are small)  scales as $\Sigma (\omega_m) \propto \omega^{1/2}_m$~\cite{millis_92,acs}
 \footnote{The actual value of $\gamma$ is somewhat different from $1/2$ as corrections to fermion-boson vertex are logarithmical, and series of these corrections change the exponent $\gamma$ to $1/2 + \epsilon$, where $\epsilon$ is positive, but small numerically~\protect{\cite{acs,morr_97}}.  Besides,  the dynamical exponent $z$ also flows exponentially from $z=2$ to a smaller value~\cite{metlitski2010quantum2}. It has been argued~\cite{lunts_2017} that in the absence of a superconducting instability, this flow eventually, at the lowest energies, brings the system into the basin of attraction of a stable fixed point with $z=1$. Our analysis is valid at energies where the dynamical exponent is still $z \approx 2$.}.
  The specific heat in the non-regularized $\gamma$-model scales as $\Lambda/\sqrt{T}$, and the one in the regularized $\gamma$-model scales as $\sqrt{T}$.  Both  are inconsistent with the specific heat of the underlying fermion-boson model $C_{afm} \propto T \log T$, as we
   show below.   The reason for the inconsistency is, however, rather banal -- the non-FL behavior, described by the $\gamma =1/2$ model, holds only in hot regions. Away from these regions
   the self-energy has a Fermi-liquid form at the smallest frequencies.  Hot fermions contribute most to superconductivity, and the $\gamma =1/2$ model of hot fermions adequately describes the interplay between non-FL and  pairing.  However, the free energy in the normal state is the combined contribution from fermions  over the whole  Fermi surface, and the one from hot fermions is proportional to the total width of the hot regions, which is small compared to the circumference of the Fermi surface boundary.

  To obtain the specific heat, we compute the free energy using Eq. (\ref{Fe_8}). We assume that $\Sigma_k$ depends on frequency and on the position on the Fermi surface, but not on $\epsilon_k$.  In this situation, one can still explicitly integrate over the dispersion in the first two terms in (\ref{Fe_8}). The result is that
  the self-energy cancels out, even if it depends on the momentum along the Fermi surface, and the free energy is given by
  Eq. (\ref{Fe_9}).
   As before, we assume that $D^{0}_q$ has Ornstein-Zernike form
   $D^{0}_q = - D_0/(({\bf q}-{\bf Q})^2 +m^2)$
   and incorporate the renormalizations from the static polarization bubble $\Pi (0)$ into $m$ and $D_0$. We evaluate the dynamical Landau damping term  $\Pi (\Omega_m) - \Pi (0) = \alpha |\Omega_m|$ right at ${\bm q} = {\bm Q}$.
   The full propagator is
   \beq
   D_q = {D_0 \over ({\bf q}-{\bf Q} )^2 +m^2 + \alpha \rvert \Omega_m \rvert }.
   \label{lll}
   \eeq
   Integrating over the component of ${\bf q}-{\bf Q}$ along the FS, we obtain
   \beq
  D_{loc} (\Omega_m) = \frac{{\bar g}^{1/2}}{ \left( \Omega_m^2 + M^2\right)^{1/4} },
   \eeq
   where
   ${\bar g}^{1/2} =
   g^2 D_0/(4 \pi v_F \sqrt{\alpha})$ and $M \sim m^2$.
   This is approximately Eq.~(\ref{dloc_1}).

   Substituting $D_q$ from (\ref{lll}) into (\ref{Fe_9}) and integrating over ${\bm q}$, we obtain
  \beq
   F_{afm} = F_{free} -\frac{3 \alpha}{2}  T^2 \sum_{n=1}^{M_B} n\log{\frac{nT}{T_0}}
   \label{Fe_28_1}
   \eeq
  where the factor of $3$ is due to summation over spin components and $T_0 \sim k^2_F/\alpha$ is a non-universal scale related to the upper cutoff of the integral over ${\bm q}$.
    Evaluating the frequency sum
  (see Appendix~\ref{sec:afm_app} for details)
  and  using (\ref{Fe_ex_2}) to relate $M_b$ and the energy cutoff $\Lambda$, we obtain
  \bea
   F_{afm} &=& -\Lambda^2 \left( N_F + \frac{3\alpha}{16 \pi^2} \log{\frac{\Lambda}{2\pi T_0 \sqrt{e}}}\right) \nonumber \\
   && + \frac{\pi^2}{3} T^2 \left (-N_F + \frac{3\alpha }{8\pi^2} \log{\frac{T}{T^*_0}}\right)
   \label{Fe_27}
   \eea
   where $T^*_0$ differs from $T_0$ by a factor $O(1)$.
  Differentiating  over $T$, we obtain
  \beq
  C_{afm} (T) = S_{afm} (T) = \frac{2\pi^2}{3} T \left(N_F  + \frac{3}{8\pi^2}
  \alpha  \log{\frac{T^*_0}{T
  e^{3/2}
  }}\right)
    \label{Fe_27_3}
  \eeq
  We see that at small $T$ the interaction contribution to specific heat is
  larger than the one from free fermions, hence $C_{afm} (T)$  scales as $T \log (T^*_0/T)$.  This behavior has been extensively discussed in the context of
  non-FL
    behavior of cuprates and heavy fermion materials (see, e.g., \onlinecite{varma} and references therein).

The prefactor $\alpha$ in (\ref{Fe_27}), (\ref{Fe_27_1})  can be expressed in terms of the effective electron-boson coupling $g^*$ and Fermi velocities at hot spots
${\bm k}_{hs}$ and ${\bm k}_{hs} +{\bm Q}$.
We define $\epsilon_{{\bm k}+ {\bm k}_{hs}} = v_x k_x + v_y k_y$, $\epsilon_{{\bm k}+ {\bm k}_{hs}+{\bm Q}} = v_x k_x - v_y k_y$ ($v^2_F = v^2_x + v^2_y$), see Fig. \ref{fig:fs_afm}.
In these notations~\cite{acs},
 \beq
 \alpha = \frac{4 g^*}{\pi v^2_F \beta}
   \label{Fe_29}
  \eeq
  where $\beta = 2v_x v_y/v^2_F$.
  Substituting into (\ref{Fe_27}), we obtain
  \beq
  C_{afm} (T) = S_{afm} (T) = \frac{2\pi^2}{3}N_F  T \left(1  + \frac{3 g^*}{2\pi^2 E_F \beta}
    \log{\frac{T^*_0}{T e^{3/2}}}\right)
    \label{Fe_27_2}
  \eeq
It is instructive to compare this  result with the specific heat  in a purely fermionic model, with and without regularization, but with the self-energy  averaged over the full Fermi surface.
  The self-energy at a Fermi point $k=k_F$, located at
  $\delta k_{\parallel} = \delta k$ from a hot spot, is~\cite{metlitski2010quantum2,wang_13}
   \beq
  \Sigma (\delta k, \omega_m) = \frac{3 g^*}{4 v_F} T \sum_{\Omega_m} \frac{{\text{sign}}(\omega_m + \Omega_m)}{\sqrt{\alpha|\Omega_m| + (\beta \delta k)^2}}.
  \label{Fe_30}
  \eeq
     At $T=0$,
  \beq
  \Sigma (\delta k, \omega_m) = \frac{3 g^*}{2\pi v_F \alpha} \left(\sqrt{\alpha|\omega_m| + (\beta \delta k)^2} - \beta|\delta k|\right) {\text {sign}} \omega_m
 \label{Fe_31}
  \eeq
  At a hot spot, $\Sigma (0, \omega_m) \propto |\omega_m|^{1/2}$, as in the $\gamma$ model with $\gamma =1/2$.
At the same time, the  self-energy, averaged over $\delta k$, scales as
$\log ({T_0/|\omega_m|})$.
  Such a self-energy emerges in the $\gamma$-model with $\gamma = 0+$ (Refs. \cite{son,son2,senthil,max_last,raghu2,ch_jept_21}) and hence proper comparison should be with this model.
   Indeed, substituting $\Sigma (\delta k, \omega_m)$ from (\ref{Fe_30}) into
   (\ref{Fee}),
   we obtain the
     free energy of the $\gamma =0+$ model:
\beq
F_{0+} = F_{free} - \frac{3 g^*}{2\pi v^2_F \beta} T^2 \sum_{n,n'} {\text{sign}} (\omega_n \omega_{n'}) \log{\frac{T^{**}_0}{|\omega_n-\omega_{n'}|}}
\label{Fe_32}
\eeq
where $T^{**}_0$ is of the same order as $T_0$.
The regularized free energy is
\beq
{\bar F}_{0+} = F_{free} - \frac{3 g^*}{2\pi v^2_F \beta} T^2 \sum_{n,n'} \left({\text{sign}} (\omega_n \omega_{n'})-1\right) \log{\frac{T^{**}_0}{|\omega_n-\omega_{n'}|}}
\label{Fe_33}
  \eeq
In  Eqs. (\ref{Fe_32}) and (\ref{Fe_33}) the summation is over $-M_f <n,n'< M_f-1$.
Evaluating the sum and relating $M_F$ to energy cutoff $\Lambda$ via (\ref{Fe_ex_1}), we obtain
\bea
F_{0+} &=& -N_F \Lambda^2 \left(1 + \frac{3 g^*}{2\pi^2 E_F \beta} \log{2}\right)  \nonumber \\
&&+ N_F \frac{3 g^*}{2 \pi E_F \beta}  \Lambda T \log{\frac{T^{**}_0}{T}} \nonumber\\
&&- \frac{\pi^2}{3} N_F T^2 \left(1 + \frac{9 g^*}{4\pi^2 E_F \beta}
 \log \left(\frac{T_1}{T} \right) \right)
\label{Fe_34}
  \eea
  and
\bea
{\bar F}_{0+} &=& -N_F \Lambda^2 \left(1 + \frac{3 g^* }{2\pi^2 E_F\beta}   \log{\frac{0.89 \Lambda}{T^{**}_0}}\right) \nonumber \\
&-& \frac{\pi^2}{3} N_F T^2 \left(1 + \frac{3 g^*}{2\pi^2 E_F \beta }
 \log \left( \frac{T^*_1}{T} \right) \right),
 \label{Fe_35}
  \eea
 where
 $T_1 \sim \Lambda$ and $T^*_1 \sim \Lambda^2/T_0$.
 Differentiating
 with respect to temperature,
 we obtain
\beq
C_{0+} (T) = \frac{2\pi^2}{3} N_F  \left(
\frac{9 g^*}{4\pi^3 E_F \beta}  \Lambda
+ T \left(1 + \frac{9 g^*}{4\pi^2 E_F \beta}
\log \left( \frac{T_1}{ T e^{3/2} } \right)\right)
\right)
 \label{Fe_34_1}
 \eeq
 and
 \beq
{\bar C}_{0+} (T)  = S_{0+} (T) = \frac{2\pi^2}{3}N_F  T \left(1  + \frac{3 g^*}{2\pi^2 E_F \beta}
\log{\frac{T^*_1}{T e^{3/2}}}\right)
    \label{Fe_35_1}
  \eeq
Comparing (\ref{Fe_27_2}) and (\ref{Fe_35_1}) we see that the prefactor for the
$T \log T$
term is the same,
  i.e., for an antiferromagnetic QCP regularization of the free energy of a purely electronic $\gamma = 0+$ model yields the same $ C(T) \sim T \log{T}$
 as from
  the bosonic term. This agrees with the analysis of Sec.~\ref{sec:1/3} of the $\gamma =1/3$ model, extended to arbitrary $0<\gamma <1$, where  we found that
  the leading $T^{1-\gamma}$ terms in ${\bar C}_\gamma (T)$ and in the full $C (T)$
   differ by a factor $1-\gamma$, which tends to one at $\gamma \to 0$.  The specific heat in the
   non-regularized
   $\gamma =0+$ model has a parasitic temperature-independent   piece that scales with $\Lambda$. The prefactor for the universal $T \log{T}$ term in (\ref{Fe_34_1}) is larger than the one in (\ref{Fe_35_1}) by the factor $3/2$ -- the same number as we found in Sec.\ref{sec:1/3}.

  Away from the critical point, $m$ is finite, and at the smallest $T$
   the $\log ({1/T})$
   dependence in (\ref{Fe_27}) is replaced by $\log ({1/m^2})$.
   The total $C_{afm} (T)$ can then be cast in the form
  \beq
  C_{afm} (T) = \frac{2\pi^2}{3}N_F  T \left(1  + \frac{3 g^*}{\pi^2 E_F \beta} \log{\frac{k_F}{m}}\right)
    \label{Fe_27_1}
  \eeq
 In this Fermi liquid regime, the self-energy, averaged along the Fermi surface, is
$ \Sigma_{av} = \lambda_{av} \omega$, where
\beq
\lambda_{av} =  \frac{3 g^*}{\pi^2 E_F \beta}  \log{\frac{k_F}{m}}
\label{Fe_36}
 \eeq
  Comparing (\ref{Fe_27_1}) and (\ref{Fe_36}), we see that in a Fermi liquid regime at a finite m,
  $C_{afm} = C_{free} (1 + \lambda_{av})$, as is expected.

\section{QCP in electron-phonon system}
\label{sec:2}

We now analyze the free energy for the case of electrons interacting with an
Einstein boson.  We use Eq. (\ref{Fe_9}) as an input and compute the bosonic contribution to the specific heat.
The propagator of an Einstein boson is
$D_q = - D_{0}/(\Omega^2_m + \omega^2_D + \Pi_q)$,
where $\omega_D$ is the bare Debye frequency and $\Pi_q$ (which incorporates the overall factor $D_0$) comes from the interaction with electrons. We set $\omega_D$ to be finite, but much smaller than the Fermi energy $E_F = v_F k_F/2$.
 We define the dimensionless coupling $\lambda$ via
 \beq
 \lambda = \frac{{\bar g}^2}{\omega^2_D},~~{\bar g}^2 = g^2
 N_F D_0
 \eeq
  where $g$ is the same as in (\ref{Fint}).
We consider temperatures {\it smaller} than $\omega_D$.  At such $T$, the contribution to the specific heat from free bosons, $C_{bos} \propto e^{-\omega_D/T}$, is exponentially small.

  For definiteness we consider the 2D case.  The form of the 2D polarization operator of free fermions at small momentum and frequency is well-known:
  \beq
  \Pi_q = 2 {\bar g}^2 \left(1 -\frac{\Omega_m}{\sqrt{\Omega^2_m + (v_F q)^2}}\right)
\label{Fe_18}
  \eeq
  We assume and then verify that
  typical $v_F q$
  are
  of order $E_F$, while
  typical $\Omega_m$ for the specific heat
   are
  of order $T$.  For such $v_F q$ and $\Omega_m$,  we can compute  $\Pi_q$ to linear order in $\Omega_m$, but need a more accurate dependence on $q$.  In 2D, the static part of $\Pi_q$ remains equal to $2g^2$ for all momenta up to $2k_F$ and drops at larger momentum.  The dynamical part changes between small $q$ and $q \sim k_F$, and for arbitrary $q<2 k_F$ is
   \beq
  -2 {\bar g}^2 \frac{|\Omega_m|}{v_F q}  \frac{2k_F}{\sqrt{4k^2_F - q^2}}.
   \eeq
  Substituting $\Pi_q$ at small $\Omega_m$  and arbitrary $q <2k_F$  into $D_q$, we obtain
\beq
D^{-1}_q = \Omega^2_m + {\bar \omega}^2_D + 2 {\bar g}^2 \frac{|\Omega_m|}{v_F q} \frac{2k_F}{\sqrt{4k^2_F -q^2}}
\label{Fe_19}
  \eeq
   where ${\bar \omega}_D = \omega_D (1-2\lambda)^{1/2}$ is the dressed Debye frequency.   The dressed ${\bar \omega}_D$ vanishes at $\lambda =1/2$ for all momenta $q < 2k_F$.
    At larger $q$, the static $\Pi_q$  decreases and ${\bar \omega}_D$ remains finite even at $\lambda =1/2$.

 Strictly speaking, the  polarization operator has to be computed using full fermionic propagators, which include the self-energy.  This does affect
 the
 static $\Pi$, which is generally different from $2g^2$ and has contributions from fermions with energies of order $E_F$,
 of the order of
 the upper cutoff $\Lambda$ in the $\gamma-$model (Ref. \cite{ch_ma}).
 To simplify the discussion,  below we keep the free-fermion result with the understanding that the actual renormalization of $\omega_D$
 likely
 differs  somewhat from $(1-2\lambda)^{1/2}$.
  The
  corrections to
  the
  Landau damping term $|\Omega_m|/v_F q$
  is of order of
   $\lambda_E$ and
  hence  are
  small.   We assume without proof that this holds even when we extend the Landau damping formula   to $q \sim k_F$.

 We show below that within the
 regime of
  validity of the Eliashberg theory,
 $\lambda_E \ll  1$,
 the last term in (\ref{Fe_19}) is small compared to the first two.  The corresponding $\gamma$-model
 then has
 $\gamma =2$.

The vanishing of the dressed Debye frequency at some finite $\lambda$ ($\lambda =1/2$ if we use free-fermion expression for static $\Pi_q$)  has been noticed before~(see e.g. \cite{Chubukov_2020b,yuz_2,kolya} and references therein), both in 2D and 3D systems.
However, in 3D, ${\bar \omega}_D$ is not flat for $q < 2k_F$, and the dressed ${\bar \omega}_D$ vanishes at a critical $\lambda$ only at $q=0$ and scales as $q^2$ at small $q$.
In this situation the full bosonic propagator has the same form as in
the
3D Ising-nematic model, and  the corresponding $\gamma$-model
has
$\gamma = 0+$, with the effective interaction
 \beq
D_{loc} (\Omega_m) = \log{\frac{\bar g}{|\Omega_m|}}
\label{Fe_20}
\eeq
  In 2D, corrections to free-fermion form of $\Pi_q$  also introduce quadratic momentum dependence of ${\bar \omega}_D$ around $q=0$, even for an  isotropic fermionic dispersion~ \cite{chubukov_93},
    such that very near QCP critical theory becomes the same as in a 2D Ising-nematic case.
Alternatively, a non-parabolic fermionic dispersion can also introduce a quadratic term in the bosonic dispersion.
However, the momentum dependence
may be weak, resulting in a
wide range around a QCP, where ${\bar \omega}_D$  can be approximated by momentum-independent
constant,
$\omega_D (1-2 \lambda)^{1/2}$.   For a system on a square lattice,  quantum Monte Carlo data show that the minimum of ${\bar \omega}_D$ is at $M =(\pi, \pi)$ (Ref. \cite{esterlis_18}).   The dispersion is flat around the minimum, and the overall variation of
${\bar \omega}_D$  with momentum is quite small.  At the minimum, ${\bar \omega}_D$  displays $(1-2 \lambda)^{1/2}$ dependence up to $\lambda  \sim 0.4$ (Ref. \cite{Chubukov_2020b}).

\begin{figure}
  \includegraphics[width=9cm]{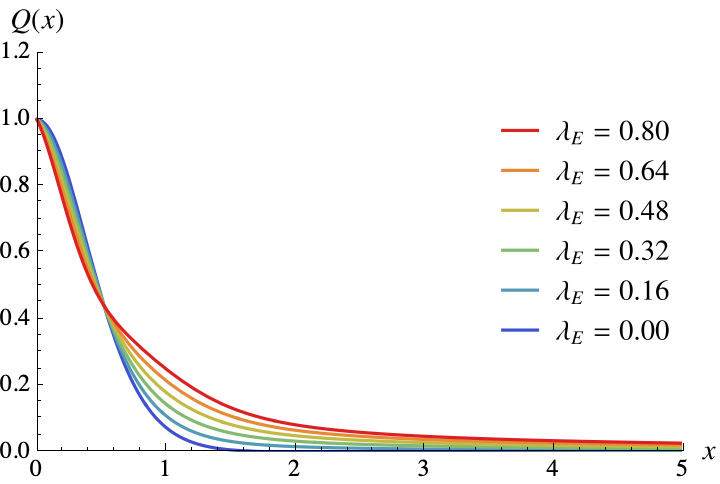}
  \caption{Function $Q(x)$, given by Eq. \protect\ref{Fe_24}, for different $\lambda_E <1$.}
  \label{fig:qx}
\end{figure}

\begin{figure}
\includegraphics[width=9cm]{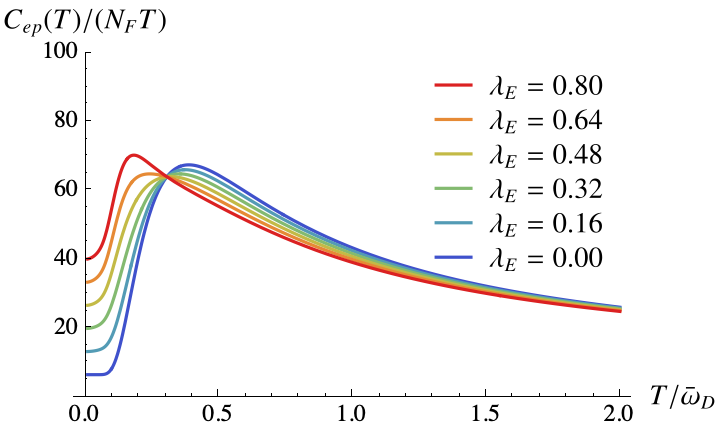}
\caption{
The ratio $C_{ep} (T)/T$, where $C_{ep} (T)$ is the specific heat of the electron-phonon system, given by  Eq.~(\ref{Fe_23_1}). We plot $C(T)/T$  a function $T/{\bar \omega}_D$, where ${\bar \omega}_D$ is the dressed Debye frequency, for different values of the  Eliashberg parameter $\lambda_E$.
  We set
$E_F = 10 \bar{g}$,
in which case $\lambda_E = 0.1 ({\bar g}/{\bar \omega}_D)^2$.
The specific heat is positive at all $T$. At $T \ll \bar{\omega}_D$, $C_{ep} (T)/T$ saturates at $2\pi^2/3 N_F
(1+\lambda/(1-2\lambda))$
at $T\gg \bar{\omega}_D$, $C_{ep} (T)/T$ asymptotically approaches
 its value for free bosons.}
\label{fig:cvep}
\end{figure}

In our analysis of the free energy we focus on the regime where the momentum dependence of ${\bar \omega}_D$ can be neglected.
 In this regime $F = F_{free} + (T/2) \sum_q \log (-{D^{-1}_q}) $,
 where
 $F_{free}= -\pi^2 T^2 N_F/3$
  is the free energy of a free Fermi gas.
 We assume and then verify that the largest contribution to the specific heat comes from the
 ${\bm q}-$independent
 term in $D_q$ and approximate $\log (-{D^{-1}_q})$
 by expanding to leading order in the Landau damping term (which we shall later show is a small correction for an $\gamma>1$)
\beq
\log (-{D^{-1}_q} ) = \log{\left(\Omega^2_m + {\bar \omega}^2_D\right)} + 2 \frac{{\bar g}^2}{v_F q} \frac{|\Omega_m|}{\Omega^2_m + {\bar \omega}^2_D} \frac{2k_F}{\sqrt{4k^2_F -q^2}}
\label{Fe_21}
\eeq
Substituting into the free energy and integrating over $|{\bf q}| < 2k_F$, we obtain
\bea
&&\frac{T}{2} \sum_q \log (-{D^{-1}_q} ) =  \frac{k^2_F}{\pi} T \sum_{n=1}^{M_b} \log{(4\pi^2 T^2 n^2 + {\bar \omega}^2_D)} \nonumber \\
&& +
\frac{{\bar g}^2 k_F}{2\pi v_F}
\sum_{n=1}^{M_b} \frac{n}{n^2 + ({\bar\omega}_D/(2\pi T))^2}
\label{Fe_22}
\eea
The first term is the free energy of a free Einstein boson with the renormalized Debye frequency ${\bar \omega}_D$, the second one is the contribution from fermion-boson interaction.
 Evaluating the frequency sums
 (see the Appendix~\ref{sec:phonon_app} for details)
 and using  the relation between $M_b$ and the
 upper energy cutoff $\Lambda$, Eq. (\ref{Fe_ex_2}), we obtain
 \bea
&&F = N_F \left(-\Lambda^2 + \frac{4 E_F \Lambda}{\pi} \log{\frac{\Lambda}{e}} + {\bar g}^2
\log{\frac{\Lambda}{{\bar \omega}_D}}\right)  \nonumber \\
&& + N_F \left[-\frac{\pi^2 T^2}{3} +  4 T E_F  \log{\left(1-e^{-\frac{{\bar \omega}_D}{T}}\right)}
 + {\bar g}^2
 f \left(\frac{{\bar \omega}_D}{2\pi T}\right) \right]
\label{Fe_22_a}
\eea
  where
  \beq
  f(x) = \log{x} - \frac{1}{2} \left(\psi (1+ix) + \psi(1-ix)\right)
   \label{Fe_23}
  \eeq
and $\psi (y)$ is di-Gamma function.
We see that the
$\Lambda$-dependent
terms in (\ref{Fe_22_a})
%,  which depend on the upper cutoff,
are independent
of
$T$,
and hence
do not contribute to
the
entropy
 and the
specific heat.
Differentiating twice
 with respect to
 temperature, we obtain the total specific heat for
the
isotropic electron-phonon system
 \beq
  C_{ep} (T) = \frac{2 \pi^2}{3} N_F \left[T + \frac{6 E_F}{\pi^2} Q \left(\frac{{\bar \omega}_D}{2\pi T}\right)\right]
  \label{Fe_23_1}
  \eeq
where
  \bea
  Q(x) = \left(\frac{\pi x}{\sinh{\pi x}}\right)^2 - \frac{\pi}{2}
   \lambda_E  x^2 \left(x \frac{d^2f}{dx^2} + 2\frac{df}{dx}\right)
  \label{Fe_24}
  \eea
  and $\lambda_E = {\bar g}^2/({\bar \omega}_D E_F)$ is the Eliashberg parameter.
 The Elishberg theory, which neglects vertex corrections, is valid  when $\lambda_E$ is small.
  In Fig.~\ref{fig:qx}, we plot  $Q(x)$ for different $\lambda_E$. We see that this function is positive for all $x$.  Accordingly,
 $C_{ep} (T)$ given by (\ref{Fe_23}) is also positive for all temperatures. We plot $C(T)$
 in Fig.~\ref{fig:cvep}.

 The limiting forms of $C_{ep} (T)$ are
 \beq
 C_{ep} (T) = \frac{2 \pi^2}{3} N_F T \left(1 + \frac{\lambda}{1-2\lambda}  \right)
  \label{Fe_25}
  \eeq
at $2\pi T \ll {\bar \omega}_D$,
 and
  \bea
 C_{ep} (T) &=& \frac{2 \pi^2}{3} N_F \left( T + \frac{6 E_F}{\pi^2} \left(1- \lambda_E \frac{{\bar \omega}_D}{4 T}\right) \right) \nonumber \\
 &=&  \frac{2 \pi^2}{3} N_F \left( T + \frac{6 E_F}{\pi^2}  - \frac{3 {\bar g}^2}{2\pi^2 T}\right)
  \label{Fe_26}
  \eea
  at $2\pi T \gg {\bar \omega}_D$, which  includes the case ${\bar \omega}_D  \to 0$ at
   finite $T$.
  In the two limits, the entropy $S_{ep} (T) = C_{ep} (T)$ at  $2\pi T \ll {\bar \omega}_D$, and
   \beq
 S_{ep} (T) = \frac{2 \pi^2}{3} N_F \left( T + \frac{6 E_F}{\pi^2} \log{\frac{T}{{\bar \omega}_D}} +
  \frac{3 {\bar g}^2}{2\pi^2 T}\right)
  \label{Fe_26_1}
  \eeq
  at  $2\pi T \gg {\bar \omega}_D$.
 Note that in this last limit the entropy
 is always
 positive. It diverges logarithmically at ${\bar \omega}_D  \to 0$ at a finite $T$.

We now take a more careful look at the  expression for the specific heat at $2\pi T \gg {\bar \omega}_D$.
The first term in the second line  in (\ref{Fe_26}) is the contribution from free fermions, the second is the contribution from
free bosons, but with an effective Debye frequency, renormalized by the interaction with fermions, and  the third term is the direct contribution from
the
electron-phonon interaction.  This  last term is negative and is the same as  the interaction contribution to the specific heat  in the regularized $\gamma$-model, Eq. (\ref{Fe_7_2}).   Without the middle term, the specific heat
 would become negative below a certain
 temperature,
  $T = (3/(2\pi^2))^{1/2} {\bar g}
  \simeq
  0.39 {\bar g}$, which
 exceeds the onset temperature for superconductivity $T_c
 \simeq
 0.18 {\bar g}$ (Ref.~\cite{combescot,Marsiglio_91,Wang2016}). Because of the middle term, however, the full $C(T)$ remains positive. This holds even at $T \to 0$, as one can see from the first line in (\ref{Fe_26}).  The key here is the condition $\lambda_E = {\bar g}^2/({\bar \omega}_D E_F) <1$, which requires one  to treat  the case of vanishing dressed Debye frequency as a double limit, in which $E_F$ tends to infinity simultaneously with ${\bar \omega}_D \to 0$ (Refs.~\cite{paper_5,zhang_22}).

The authors of (\cite{yuz_1,yuz_2}) argued that the negative prefactor for the $1/T$ term in Eq.~(\ref{Fe_7_2}) indicates that the normal state becomes unstable below a certain $T$ despite that the total $C_{ep} (T)$ is positive. 
Their argument is that the $T-$independent term in (\ref{Fe_26}), which renders the total $C_{ep} (T)$ positive, is the contribution from free bosons and as such does not affect the electrons.  
Our counter-argument is that both positive and negative parts of $C_{el} (T)$  come from the term in the free energy $(T/2) \sum_q \log{[-D^{-1}_q]}$, once one expands it in the dynamical part of $\Pi_q$: 
the positive contribution is the the zeroth order term and the negative $1/T$ contribution comes from the first order in the expansion.
In our view, this shows that both terms should be treated on equal footings.
Besides, despite the fact that the leading $T-$ independent part of $C_{el} (T)$ has the same form as the specific heat of a free massless boson, this term does depend on fermion-boson interaction 
as the latter renormalizes the bare $\omega_D$ into ${\bar \omega}_D = \omega_D \sqrt{1-2 \lambda}$.  
For $\lambda \approx 1/2$, $\omega_D = {\bar g}/\sqrt{\lambda} \approx \sqrt{2} {\bar g}$ is comparable to ${\bar g}$, and
without interaction-driven renormalization of $\omega_D$ into ${\bar \omega}_D$ the specific heat of free bosons would be exponentially small at $T \leq {\bar g}$.  In this respect,
the fermion-boson coupling gives rise to two effects: it generates a negative $T$-dependent contribution to $C_{ep}$, and simultaneously gives rise to a much larger, positive $T-$independent contribution.

\begin{figure}
\includegraphics[width=9cm]{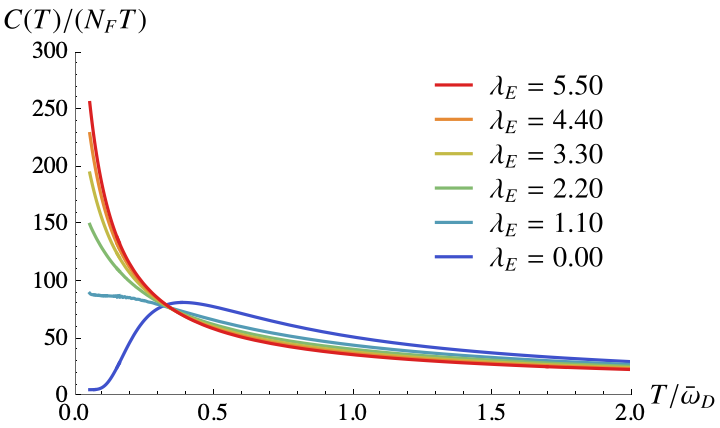}
\caption{
Normalized specific heat of the electron-phonon system, $C_{ep} (T)/T$,  obtained by using the full expression for the free energy $F=F_{free}+(T/2)\sum_q \log ( -D_q^{-1})$, without expanding $\log ( -D_q^{-1})$ in the Landau damping. We used the same parameters as in  Fig.~\ref{fig:cvep}. The specific heat remains positive for all values of $\lambda_E$.}
\label{fig:cvep_full}
\end{figure}

For completeness, we also compute the specific heat within our model for larger $\lambda_E$,  using the full formula
for $\log{D^{-1}_q}$ rather than expanding in the Landau damping
in Eq. (\ref{Fe_21}).  We find that the specific heat is positive for all $\lambda_E$. We plot $C_{ep}/T$ in Fig.~\ref{fig:cvep_full}.   
This result is of limited validity, however, as in the free energy we didn't include higher-order terms in the skeleton expansion in $F_{int}$.  These terms are of higher order in  $\lambda_E$, when $\lambda_E$ is small, but are
not small when $\lambda_E >1$. Still, we emphasize that within the model we used here, $C_{ep}(T)$ is positive for all $T$.

\subsection{Physical origin of the regularization of $F_\gamma$}
\label{sec:new}

We now argue that the interaction-driven renormalization of $\omega_D$ is related to the issue of the regularization of $F_\gamma$ in the $\gamma$-model. To relate the two, we recall that $i T \sum_k \Sigma_k G_k$, which is the interaction part of $F_\gamma$,  can be
 re-expressed as $(T/2) \sum_q \Pi_q D_q$ (see Eq. (\ref{eqn_5}), where
 $q = ({\bf q}, \Omega_m)$
 and $\Pi_q =2 g^2 T \sum_k G_k
 G_{k+q}$.
 In the analysis above, we computed this last term neglecting in $D_q$ the dynamical part of $\Pi_q$, which also depends on momentum ${\bf q}$.
   Without
   this
   term, $D_q$ depends only on frequency, and the momentum integration
   involves only $\Pi_q$.  The double integral over ${\bf q}$ and ${\bf k}$  can be transformed into the
    integration over the two fermionic momenta ${\bf k}$ and ${\bf k}+{\bf q}$ and then into the integration over the two dispersions $\epsilon_k$ and $\epsilon_{k+q}$. Each integral is proportional to ${\text {sign}} \omega_m$, where $\omega_m$ is the Matsubara frequency in the corresponding Green's function, hence the momentum integration gives rise to the factor ${\text {sign}} (\omega_m \omega_{m'})$, where $\omega_m -\omega_{m'} = \Omega_m$.  This is the same factor as in the second term in Eq. (\ref{Fe_2}) for the free energy $F_\gamma$ of the non-regularized $\gamma$-model. The same holds for the interaction term in the free energy in the $\gamma$ model:
    ${\text {sign}} (\omega_m \omega_{m'})$ in the interaction term in (\ref{Fe_2}) has been obtained by integrating independently over two fermionic dispersions: one of $G_{k-q}$ in Eq. (\ref{El_s}) and the other of $G_k$ in $i T \sum_k \Sigma_k G_k$.
      Using now ${\text {sign}} (\omega_m \omega_{m'}) =1 + ({\text {sign}} (\omega_m \omega_{m'})-1)$,
        we immediately see that the first and second terms correspond to contributions from the static  and dynamical parts of $\Pi_q$, respectively.

        Hence,  the static part of $\Pi_q$
        accounts
        for the renormalization of the bare
        $\omega_D$ into ${\bar \omega}_D$, which vanishes at the QCP.
    In the underlying fermion-boson model, $\Pi_q$ is the full polarization operator, with static and dynamics parts, and the renormalization  $\omega_D \to {\bar \omega}_D$  must be
    taken
    into consideration.
     This implies that
        $(T/2) \sum_q \Pi_q D_q$  and $i T \sum_k \Sigma_k G_k$ have to be computed without adding
        counter
        terms, and both
        depend on the
        upper
                cutoff. Like we demonstrated,  the two terms cancel out in the full free energy $F$. The latter is expressed in terms of $D^{-1}_q$, which contains the dressed ${\bar \omega}_D$ and the dynamical part of $\Pi_q$.

         Then even when the renormalization of the bosonic mass does depend on the cutoff (e.g., in the case of a lattice dispersion), the free energy is expressed via the fully dressed mass, which vanishes at a QCP.

 The $\gamma-$model is constructed differently. In this model, the renormalization of $\omega_D$ into
         ${\bar \omega}_D$ is already absorbed into $D_{loc} (q)$, which, by construction, depends on the dressed
          ${\bar \omega}_D$. Hence,  the terms
          which renormalize $\omega_D$
          must be excluded to avoid
          double
          counting.  The way to do this is to eliminate the contribution from
          the static part of
          $\Pi_q$ by replacing
          ${\text {sign}} (\omega_m \omega_{m'})$ by ${\text {sign}} (\omega_m \omega_{m'})-1$.  This is precisely the
          counter
          term, which the authors of \cite{yuz_1,yuz_2,yuz_3} suggested to add to regularize the free energy of the $\gamma$-model.

          The same reasoning holds for other values of $\gamma$. In each $\gamma-$model, one has to
          subtract the renormalization of the bosonic mass to avoid double counting.  This is achieved by
           the same substitution
           ${\text {sign}} (\omega_m \omega_{m'})$ by ${\text {sign}} (\omega_m \omega_{m'})-1$ in Eq. (\ref{Fe_2}).

\section{Extension to $\gamma < 2$}
\label{sec_ext}

It is instructive to verify how the $T-$independent and the $1/T$ term in Eq. (\ref{Fe_26}) evolve if we add a  momentum-dependent term to
 the bosonic propagator  $D_q$ in (\ref{Fe_19}) and gradually change the exponent $\gamma$ in the corresponding $\gamma$-model to $\gamma <2$.  A way to do this phenomenologically is to consider  a fermion-boson model with the
bosonic propagator
\beq
D^{-1}_q = \Omega^2_m + (c q)^{2a} + {\bar \omega_D}^2 +  2 {\bar g}^2 \frac{|\Omega_m|}{v_F q}
\label{Fe_19_1}
  \eeq
 with $a >1$.
 We assume that the $q^{2a}$ term comes from fermions with energies of order $E_F$,  and  set the prefactor $c$ to be of order $E^{1/a}_F/k_F$. As before, we consider the double limit in which ${\bar \omega}_D$ tends to zero and simultaneously $E_F$ tends to infinity.

 We verified that the leading contribution to the fermionic self-energy $\Sigma (\omega_m)$ comes from the first two terms in (\ref{Fe_19_1}), while the Landau damping term accounts for a negative correction.  Specifically,
   \beq
   \Sigma (\omega_m) \propto |\omega_m|^{1/a-1} \left(1 - \left(\frac{T_a}{|\omega_m|} \right)^{\frac{a+1}{a}}\right)
   \label{Fe_28_4}
  \eeq
  where
  \beq
  T_a \sim {\bar g} \left(\frac{{\bar g}}{E_F}\right)^{\frac{a-1}{a+1}}
   \label{Fe_28_5}
   \eeq
  For
  $E_F \to \infty$,  $T_a$ tends to zero, hence $T_a/|\omega_m|$ is vanishingly small for all $\omega_m$.  Comparing (\ref{Fe_28_4}) with  $\Sigma (\omega_m) \propto |\omega_m|^{1-\gamma}$ in the $\gamma-$model, we find $\gamma =2-1/a$. This exponent ranges between $1$ and $2$, when $a$ ranges between $1$ and infinity.  At $a = 1+0$, a more accurate analysis shows that $\Sigma (\omega_m) \propto \log{|\omega_m|}$.

The free energy and the specific heat can be obtained in the same way as above.
For brevity,
we skip the details of
the
calculations and just list the results. We also neglect
the
free-fermion part of the specific heat and label the specific heat due to fermion-boson interaction as $C_{int} (T)$.
Up to a positive  overall factor,
\beq
 C_{int} (T) \propto T^{\frac{2}{a}} \left( 1 - \left(\frac{T_a}{T}\right)^{\frac{a+1}{a}} +...\right)
 \label{Fe_28_6}
 \eeq
 where dots stand for higher-order terms in the expansion in $T_a/T$.
 The positive term in (\ref{Fe_28_6}) comes from the $\Omega^2_m$ and $(cq)^{2a}$ terms in the bosonic ptopagator, and
   the negative term comes from the Landau damping term  in  (\ref{Fe_19_1}).
    This negative term is vanishingly small as $T_a$ tends to zero when $a >1$.
     The exponent $1/a$ equals to $2-\gamma$,  hence $C_{int} (T) \propto
     T^{2(2-\gamma)}$. For $\gamma =2$,
      $C_{int} (T)$ becomes temperature independent. This is consistent with the result that we obtained
      in the previous Section.

 \subsection{Extension to $1/2 <a<1$}

 For completeness, we also present the results for smaller values of the exponent $a$: $1/2 < a <1$. The condition $a >1/2$ is required for ultraviolet convergence.

Evaluating the fermionic self-energy, we now obtain
 \beq
   \Sigma (\omega_m) \propto |\omega_m|^{\frac{2}{2a+1}} \left(1 - \left(\frac{|\omega_m|}{T_a} \right)^{\frac{a+1}{a}}\right)
   \label{Fe_27_4}
  \eeq
  where $T_a$ is the same as in  (\ref{Fe_28_5}).
  The dominant contribution to the self-energy now comes from the Landau damping term and from the $(c q)^{2a}$ term in $D_q$ in (\ref{Fe_19_1}),  while the $\Omega^2_m$ term accounts for a negative correction.
  Because
  $T_a$ now tends to infinity at $E_F \to 0$,
   the second term in (\ref{Fe_27_1}) is vanishingly small for all $\omega_m$.  Associating the exponent $2/(2a+1)$ with $1-\gamma$, we find that for $ a <1$, $\gamma = (2a-1)/(2a+1)$.

 For the specific heat we find
 \beq
 C_{int} (T) \propto T^{\frac{2}{2a+1}} \left( 1 - \left(\frac{T}{T_a}\right)^{\frac{a+1}{a}}+ ...\right)
 \label{Fe_27_6}
 \eeq
 The positive contribution to $C(T)$ now comes from  the Landau damping term and the $(cq)^{2a}$ term in (\ref{Fe_19_1}), while the negative contribution comes  from the $\Omega^2_m$ term.  The dots stand for terms with higher powers of $T/T_a$.
 Because for $a <1$,  $T_a$  tends to infinity at $E_F \to \infty$, the negative term is vanishingly small at any $T$. As a result $C_{int} (T)$ is again positive.  Using the relation $\gamma = (2a-1)/(2a+1)$, valid for $a <1$, we find that  $C_{int} (T) \propto T^{1-\gamma}$.  This agrees with the results in Sec. (\ref{sec:1/3}). At $a \to 1/2$, a more accurate analysis  yields $C_{int} (T) \propto T \log T$, as in Sec. (\ref{sec:1/2}).

There is a discontinuity in $\gamma$ at $a =1$, i.e.,
 the model with $a = 1+0$ corresponds to
$\gamma =1$, and the one with  $a = 1-0$ corresponds to $\gamma =1/3$.
 This  is the consequence of discontinuity of $T_a$ at $a =1$ and $E_F \to \infty$: $T_a$  tends to zero at $a >1$, is of order ${\bar g}$ at $a =1$, and tends to infinity at $a <1$. Right at $a =1$, the frequency dependence of the self-energy and the temperature dependence of the specific heat undergo a
crossover  from $\Sigma (\omega_m)  \propto |\omega_m|^{2/3}$ and $C_{int} (T) \sim T^{2/3}$ at $\omega_m, T \ll {\bar g}$ to $\Sigma (\omega_m)  \propto \log{|\omega_m|}$  and $C_{int} (T) \sim T^2$  at $\omega_m, T \gg {\bar g}$ (modulo logarithms).  In both cases the specific heat is positive.  The low-temperature behavior of the model with $a=1$ is the same as in the $\gamma =1/3$ model.

\section{Conclusions}
\label{sec:concl}

In this paper, we analyzed
the  free energy and specific heat for a system of fermions interacting with
nearly gapless bosons near a QCP in a metal.
 The effective low-energy model for quantum-critical fermions is the one in which bosons are integrated out, and
the
fermions are interacting
via
an effective, purely dynamical interaction $V (\Omega_m) \propto 1/|\Omega_m|^\gamma$. This $\gamma-$model is adequate for the description of
non-FL
behavior and pairing near a QCP, and the competition  between tendencies towards
non-FL
and pairing. This physics is fully determined by low-energy fermions and is independent on the upper energy cutoff in the theory, $\Lambda$.  The condensation energy, associated with pairing, is also independent
of
$\Lambda$.  At the same time
within the $\gamma$-model,
the free energy in the normal state
    does depend on $\Lambda$. Furthermore, the dependence on $\Lambda$ extends to temperature-dependent terms in
    the free energy. As a result, the specific heat in the $\gamma-$model  also depends on the cutoff.
    In recent papers~\cite{yuz_2,yuz_3},
    the authors argued that the dependence on $\Lambda$ is a
    spurious
    one and has to be eliminated by proper regularization.  They added a term to the free energy, which cancels out cutoff dependence of the free energy.  However, the regularized specific heat turns out to be negative for $\gamma \geq 2$, considered in \cite{yuz_2,yuz_3}.
    Some of us and others~\cite{zhang_22}
    argued that
    the
    specific heat becomes negative at small enough $T$ already at $\gamma >1$.

   We analyzed the specific heat near
   the
   QCP by returning back to the underlying fermion-boson model
     and collecting contributions to the free energy from fermions, bosons, and their interaction.
    This allowed us to obtain the full expression for the specific heat and compare
    it
    with the regularized specific heat in fermions-only $\gamma-$model.

  Our key result is that
  the specific heat
  in the full fermion-boson model
  is independent on the cutoff and is positive all the way up to a QCP.
  This holds within the Eliashberg theory, which we used in the calculations, in the parameter range where the theory is rigorously justified, i.e., the dimensionless Eliashberg parameter $\lambda_E$, which measures the strength of vertex corrections, is small (number-wise, $C(T)$ remains positive even when
   $\lambda_E =O(1)$).

We  considered three cases, all in 2D: Ising-nematic QCP,  antiferromagnetic QCP, and QCP for electrons interacting with Einstein phononons.  For the first case, the exponent in the purely electronic model is $\gamma =1/3$.
For the second it is $\gamma =1/2$ for
fermions near the hot spots,
but is reduced to $\gamma =0+$ in the effective model with the interaction averaged over the Fermi surface.
For
electron-phonon case,
the effective fermion-only model
has  $\gamma =2$.

For the two cases with $\gamma <1$,
where  the specific heat in the regularized $\gamma-$model is positive, we found that the regularization and the effect of keeping the bosonic piece in the free energy
  is largely the same thing. Specifically, the regularized specific heat has correct temperature dependence
 ($T^{2/3}$ for the Isng-nematic case,
and
 $T \log T$
 for the AFM case), and the prefactor differs from the correct one only by a numerical factor, which, moreover,
 is
 equal to one in the AFM case.

In
 the electron-phonon case ($\gamma =2$) the modified electronic specific heat reproduces the temperature dependence of the actual $C(T)$.
 However, $C(T)$ has
  an additional temperature-independent piece, which also comes from
  the
  electron-phonon interaction.
   Both terms originate from $\log{D_q}$, and the temperature-independent term is the leading one.
  Because of this,
    the actual $C(T)$ is positive for all values of the dressed Debye frequency, i.e., at any distance from
  the
  QCP.
   The same holds for other models, whose fermionic part is described by the $\gamma$ model with $\gamma >1$.
   We believe that this result implies that the normal state of a critical fermion-boson model
   remains stable at all $T$, as long as one neglects the pairing instability.
   In this, our conclusions differ from the ones in Refs. \cite{yuz_2,yuz_3}.

\section{Acknowledgement}

\paragraph*{\bf{Acknowledgment}}~~~We thank Ar. Abanov, B. Altshuler, A. Klein,  A. Levchenko, D. Maslov, J. Schmalian, G. Torroba, Y. Wang, Y. Wu, and E. Yuzbashyan  for fruitful discussions. This project was supported by the US-Israel Binational Science Foundation (BSF).
 The work by A.V.C. was supported by  U.S. Department of Energy, Office of Science, Basic Energy Sciences, under Award No. DE-SC0014402.
 E.B. was supported by the European Research Council (ERC) under grant HQMAT (Grant Agreement No. 817799).

\appendix

\section{Details about Ising-nematic case}

\subsection{Self-energy of an electron at a finite $T$}

The one-loop self-energy of an electron is given by
\begin{equation}
i\Sigma_k=-g^*T\sum_{\Omega_{m}}\int\frac{d^{2}\bm{q}}{(2\pi)^{2}}\frac{1}{i\tilde{\Sigma}_{k+q}-\epsilon_{\bm{k}+\bm{q}}}
\frac{D_0}{q^{2}+m^{2}+\alpha\frac{\rvert\Omega_{m}\rvert}{\rvert\bm{q}\rvert}},\label{eq:sigma}
\end{equation}
where $\Sigma_k = \Sigma({\bf k}, \omega_m)$ and the notations are the same as in the main text:
$\tilde{\Sigma}_k\equiv\omega_{m}+\Sigma_k$ and
$\alpha=g^{*}k_{F}/(\pi v_{F}^{2})$, where
$g^*$  is the effective fermion-boson coupling.

At $T=0$, the sum is replaced by $T\sum_{\Omega_{n}}=(1/2\pi)\int d\Omega_{n}$.
 The leading term in $\Sigma_k$ is obtained by
 factorizing the
momentum integration along and transverse to the Fermi surface (see Fig.~\ref{fig:FS_IN} of the main text).
This leading term depends only on frequency, i.e., the self-energy is local. At a QCP,
\begin{equation}
\Sigma_k=\frac{3}{2}\bar{g}^{1/3}\rvert\omega_{m}\rvert^{2/3}\text{sgn}(\omega_{m}),\label{eq:self_0T}
\end{equation}
where $\bar{g}$ - the coupling constant of the corresponding fermionic $\gamma =1/3$ model is
\begin{equation}
\bar{g}=\frac{1}{3^{9/2}}\left(\frac{v_{F}}{k_{F}}\right)^{3}\alpha^{2}=\frac{1}{3^{9/2}\pi^{2}}\frac{(g^{*})^{2}}{E_{F}}.
\end{equation}
The factorization of momentum integration is valid as long as
 typical  fermionic momenta $q^{\text{typ}}_{f}\sim\text{max}(\tilde{\Sigma}(\omega_m ),\epsilon_{\bm{k}})/v_{F}$ (same as
 typical momenta transverse to the Fermi surface $q^{\text{typ}}_{\perp}$) is much smaller than typical bosonic momentum
 $q^{\text{typ}}_{b} \sim (\alpha |\Omega_m|)^{1/3}$ (same as typical momenta along the Fermi surface $q^{\text{typ}}_{\parallel}$).  The comparison of the two scales shows that the factorization is valid in the whole range where $\Sigma_k > \omega_m$ and at larger frequencies holds up to $\omega_{max} \sim (g^* E_F)^{1/2} \sim {\bar g}^{1/4} E_F^{3/4}$.

At a finite temperature, there are two types of bosonic fluctuations:
the thermal one with $\Omega_{m}=0$ and the quantum one with $\Omega_{m}\neq0$.
This splits the self-energy into two parts~\cite{acs,DellAnna2006,*metzner_new,punk,torroba_1,*torroba_2,avi}
\begin{equation}
\Sigma_k(k)=\Sigma^{th}_k+\Sigma^{q}_k,
\end{equation}
where, we remind, $\Sigma_k = \Sigma ({\bf k}, \omega_m)$. We have
\begin{equation}
i\Sigma^{th} ({\bf k}, \omega_m) =-g^{*}T\int\frac{d^{2}\bm{q}}{(2\pi)^{2}}\frac{1}{i\tilde{\Sigma}(\bm{k}+\bm{q},\omega_{m})-\epsilon_{\bm{k}+\bm{q}}}
\frac{1}{|{\bm q}|^{2}+m^{2}},\label{eq:Sigma_th}
\end{equation}
and
\begin{equation}
i\Sigma^{q} ({\bf k}, \omega_m) =-g^{*}T\sum_{\Omega_{n}\neq0}\int\frac{d^{2}\bm{q}}{(2\pi)^{2}}\frac{1}{i\tilde{\Sigma}({\bm k}+{\bm q}, \omega_m + \Omega_m)-\epsilon_{\bm{k}+\bm{q}}}\frac{1}{|{\bm q}|^{2}+m^{2}+\alpha\frac{\rvert\Omega_{n}\rvert}{\rvert\bm{q}\rvert}}.\label{eq:Sigma_q}
\end{equation}

Below we consider the two components of the self-energy separately.
We assume for simplicity that the bosonic mass $m$ acquires some weak temperature dependence via mode-mode coupling and
 cut $\log{m}$ singularity in the formulas  below by $\log{T}$ (for the analysis of $\Sigma_k$  for $T-$independent mass  see~\cite{avi}.  In this approximation,
   the quantum self-energy $\Sigma^q$ can  still be  computed by factorizing the momentum integration and remain local~\cite{avi,guo_22}.
   The result is
   \begin{align}
\Sigma^{q}(\omega_{m}) & =\pi T\sum_{\Omega_{n}\neq0}\left(\frac{\bar{g}}{\rvert\Omega_{n}\rvert}\right)^{1/3}\text{sgn}(\omega_{m}+\Omega_{n})\nonumber \\
 & =\bar{g}^{1/3}(2\pi T)^{2/3}H_{1/3}(m),
\end{align}
where $H_{\gamma}(m)=\sum_{n=1}^{m}1/n^{\gamma}$ is the Harmonic number. At frequencies $\omega_{m}\gg T$, one can use the expansion of a Harmonic number at large $m$:
$H_{1/3}(m)\simeq3/2 (m+1/2)^{2/3} +  \zeta (1/3) + ...$
 and obtain
\begin{equation}
\Sigma^{q}(\omega_{m})\simeq\frac{3}{2}\bar{g}^{1/3} \text{sgn}(\omega_{m}) \left( \rvert \omega_{m}\rvert^{2/3} + {2\over 3} \zeta(1/3)(2\pi T)^{2/3} +  ... \right),\label{eq:self_q}
\end{equation}
This formula is valid up to the same $\omega_{\text{max}}$ as at $T=0$.

  For thermal self-energy,  momentum integration can be factorized only in a particular parameter range, which we identify below. Outside this range, the leading contribution to $\Sigma^{th}_k$ in  (\ref{eq:Sigma_th}) is obtained by integrating over both momentum components in the bosonic propagator.

Below we consider separately  parameter ranges where $\Sigma^{th}_k$ is local and where it is not.

\subsection{Local self-energy: $\Sigma_k \equiv \Sigma(\omega_m)$}

In this section, we consider the situation when the momentum integration in Eq.~(\ref{eq:Sigma_th})
can be factorized. The factorization implies that for the same frequency,  typical fermionic momentum (the one
 transverse to the Fermi surface) is much smaller than typical bosonic  momentum connecting points on  the Fermi surface.
Typical  fermionic momentum is
$q^{\text{typ}}_{f}\sim\text{max}(\tilde{\Sigma}(\omega_m ),\epsilon_{\bm{k}})/v_{F}$, while typical bosonic momentum is
$q^{\text{typ}}_{b}\sim m$. Factorization is justified when  $q^{\text{typ}}_{f}\ll q^{\text{typ}}_{b}$.   Under this condition
\begin{equation}
i\Sigma^{th}(\omega_{m}) = - \frac{g^{*}T}{4\pi^{2}}\int_{-\Lambda_q}^{\Lambda_q}\frac{dq_{\bot}}{i\tilde{\Sigma}(\omega_{n})-
\epsilon_{\bm{k}}-v_{F}q_{\bot}}\int_{-\Lambda_q}^{\Lambda_q}\frac{dq_{\parallel}}{q_{\parallel}^{2}+m^{2}}.
\end{equation}
where $\Lambda_q \sim k_F$ is the upper cutoff of momentum integration. Assuming both $q_f$ and $q_b$  are far smaller than $\Lambda_q$, one can set $\Lambda_q \to \infty$. Momentum integration then can be done explicitly, and the result is
\begin{equation}
\Sigma^{th}(\omega_{m})=\frac{g^{*}T}{4mv_{F}}\text{sgn}(\omega_{m})\equiv\pi T\left(\frac{\bar{g}}{ M }\right)^{1/3}\text{sgn}(\omega_{m}).\label{eq:thermal_self}
\end{equation}
where
\begin{align}
M & =\frac{64}{3^{9/2}}\frac{m^{3}}{\alpha}.
\end{align}

The total self-energy $\Sigma(k)=\Sigma^{th}(k)+\Sigma^{q}(k)$ is
\begin{equation}
\Sigma(\omega_{m})\simeq\left[\pi T\left(\frac{\bar{g}}{ M }\right)^{1/3}+\frac{3}{2}\bar{g}^{1/3}\rvert\omega_{m}\rvert^{2/3}\right]\text{sgn}(\omega_{m}),\label{eq:self_loc}
\end{equation}
The two terms become comparable at
\begin{equation}
\omega_{{\text cross}}(T)\sim\frac{T^{3/2}}{ M^{1/2}}.
\end{equation}
Thermal self-energy is larger at
$\omega_{m}<\omega_{{\text cross}}(T)$.

\begin{figure}
\centering
\includegraphics[scale=0.7]{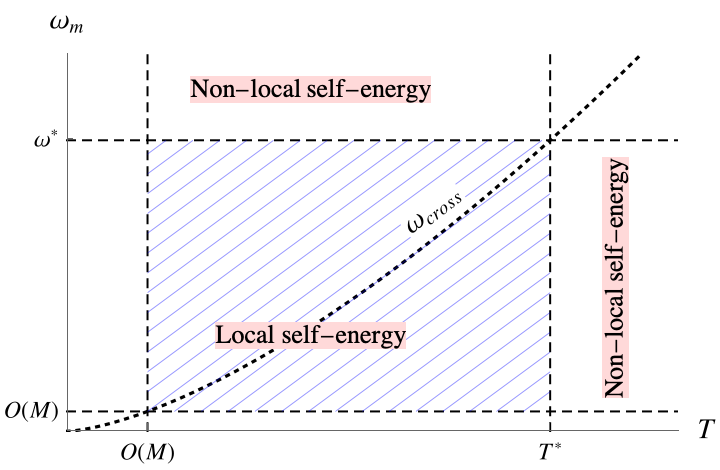}
\caption{Parameter range where the self-energy given by Eq.~(\ref{eq:self_loc}) is local, i.e., momentum-independent (marked by ``Local'' in the plot). }
\label{fig:cond_loc}
\end{figure}

Eq. (\ref{eq:self_loc}) is valid when  $q^{\text{typ}}_{f}\ll q^{\text{typ}}_{b}$. i.e., when
\begin{equation}
\tilde{\Sigma}(\omega_{m})/v_{F}\ll m.\label{eq:condition}
\end{equation}
At $\omega_{m} <\omega_{{\text cross}}$,  $\Sigma^{th}> \Sigma^{q}$, and Eq. (\ref{eq:condition})
 sets the condition on temperature
\begin{equation}
 M < T <
 T^* \sim\bar{g}\left(\frac{M}{\bar{g}}\right)^{2/3}\left(\frac{E_{F}}{\bar{g}}\right)^{1/2}.\label{eq:cond1_loc}
\end{equation}
At $\omega_{m} > \omega_{{\text cross}}$, $\Sigma^{th} < \Sigma^{q}$, and Eq. (\ref{eq:condition}) sets the condition on frequency
\begin{equation}
\omega_{{\text cross}} < \omega_{m} <
\omega^*
\sim\bar{g}\left(\frac{M}{\bar{g}}\right)^{1/2}\left(\frac{E_{F}}{\bar{g}}\right)^{3/4}.\label{eq:cond2_loc}
\end{equation}
  One can check that self-consistency condition $\omega^* > \omega_{{\text cross}}$ leads to  the  same condition on $T$ as Eq. (\ref{eq:cond1_loc}).  Then, when  Eq. (\ref{eq:cond1_loc}) is satisfied,  factorization of momentum integration is valid for all frequencies up to $\omega^*$.  We illustrate this in  Fig.~\ref{fig:cond_loc}.

We emphasize that the $T$ range in Eq. (\ref{eq:cond1_loc}) does exists at small but finite $M$ simply because $M^{2/3} > M$, but collapses at a QCP, where $M =0$.  In other words, factorization of momentum integration in the integral for $\Sigma^{th}$ holds only  away from a QCP.

There is one more condition. We assumed above that $\Sigma_{k} \gg \omega_m$. A simple analysis shows that this condition is   satisfied at arbitrary ratio of  $\Sigma^{th}$ and $\Sigma^{q}$  when $M < \bar{g}^{5/2}/E_F^{3/2}$. This relation obviously holds for small $M$.

\subsection{Non-local self-energy}

At temperatures above $T^{*}$, the condition $q_{\parallel}^{typ} \gg q_{\bot}^{typ}$ in the integral for $\Sigma^{th}$
is not satisfied. The integration over $q_{\parallel}$  in (\ref{eq:Sigma_th}) can be done explicitly:
\begin{equation}
\int_{-\Lambda_{q}}^{\Lambda_{q}}\frac{dq_{\parallel}}{q_{\parallel}^{2}+q_{\bot}^{2}+m^{2}} \approx \frac{\pi}{\sqrt{q_{\bot}^{2}+m^{2}}}.
\end{equation}
Then
\begin{equation}
i\Sigma^{th}(k)=-\frac{g^{2}AT}{4\pi}\int_{-\Lambda_q}^{\Lambda_q}\frac{dq_{\bot}}{i\tilde{\Sigma}(\omega_{m},\bm{k}+\bm{q})-\epsilon_{\bm{k}}-v_{F}q_{\bot}}\frac{1}{\sqrt{q_{\bot}^{2}+m^{2}}}.
\end{equation}
One can verify (see below) that at $T \gg  T^*$,
the leading term in this integral  is obtained by
ignoring the $\bm{q}$ dependence in the ferminonic
propagator and pulling it out of the integral, i.e., by approximating
\begin{align}
\int_{-\Lambda}^{\Lambda}\frac{dq_{\bot}}{i\tilde{\Sigma}(k+q)-\epsilon_{\bm{k}}-v_{F}q_{\bot}}\frac{1}{\sqrt{q_{\bot}^{2}+m^{2}}} & \simeq\frac{2\log\left(\frac{1}{m}\right)}{i\tilde{\Sigma}(k)-\epsilon_{\bm{k}}}.
\label{bb}
\end{align}
This leads to an algebraic relation
\begin{equation}
\Sigma^{th}(k)=\frac{B}{\left(\Sigma^{th}(k)+\tilde{\Sigma}^{q}(k)\right) + i\epsilon_{\bm{k}}},\label{eq:thermal}
\end{equation}
where $\tilde{\Sigma}^{q}(k)=\Sigma^{q}(k)+\omega_{m}$, and
\begin{equation}
B=\frac{g^{*}T}{2\pi}\log\left(\frac{1}{m}\right).
\end{equation}
Eq. (\ref{eq:thermal}), viewed as quadratic equation on $\Sigma^{th} (k)$,  has two solutions. The physical one must satify
the boundary condition $\Sigma^{th}=0$ at $B=0$. This
selects out the solution
\begin{equation}
\Sigma^{th}(k)=-\frac{\tilde{\Sigma}^{q}(\omega_{n})+i\epsilon_{\bm{k}}}{2}+\text{sgn}(\omega_{n})
\sqrt{\frac{\left(\tilde{\Sigma}^{q}(\omega_{n})+i\epsilon_{\bm{k}}\right)^{2}}{4}+B}.\label{eq:self_nonl}
\end{equation}
We remind that we define $\sqrt{z}$ with a
 branch cut along the negative real axis of the complex
variable $z$. One can verify that upon $\omega_{m}\leftrightarrow-\omega_{m}$
and $\epsilon_{\bm{k}}\leftrightarrow-\epsilon_{\bm{k}}$, $\Sigma^{th}(k)$
transforms as
\begin{align}
\text{Re}\Sigma^{th}(\omega_{m},\epsilon_{\bm{k}}) & =-\text{Re}\Sigma^{th}(-\omega_{m},\epsilon_{\bm{k}})=+\text{Re}\Sigma^{th}(\omega_{m},-\epsilon_{\bm{k}}),\\
\text{Im}\Sigma^{th}(\omega_{m},\epsilon_{\bm{k}}) & =+\text{Im}\Sigma^{th}(-\omega_{m},\epsilon_{\bm{k}})=-\text{Im}\Sigma^{th}(\omega_{m},-\epsilon_{\bm{k}}).
\label{bbbb}
\end{align}
 When $\Sigma^{th} (k) > \tilde{\Sigma}^{q}$,  $\Sigma^{th}(k) \approx \sqrt{B} \text{sgn}(\omega_{n})$.

Eq. (\ref{eq:self_nonl}) has been obtained in~\cite{avi} for $\epsilon_k =0$. We will be chiefly interested in the consequences of the dependence of $\Sigma^{th}(k)$ on $\epsilon_k$.

\begin{figure}
\centering
\includegraphics[scale=0.7]{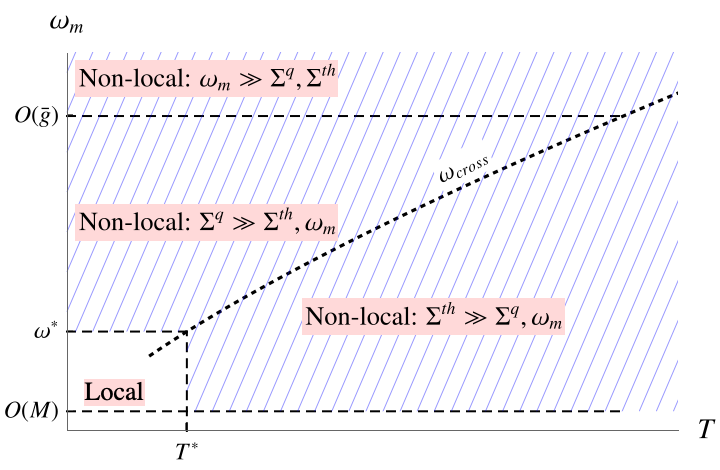}
\caption{Parameter range where  the self-energy $\Sigma^{th}(k)$  is non-local  (marked by ``Non-local'' in the plot).
}
\label{fig:cond_nonl}
\end{figure}

The total self-energy is given by $\Sigma(k)=\Sigma^{th}(k)+\Sigma^{q}(k)$,
with the quantum part given by Eq. (\ref{eq:self_q}). Expanding the
self-energy to linear order in $\epsilon_{\bm{k}}$ we find
\begin{equation}
\Sigma(\omega_{m},\epsilon_{\bm{k}})\simeq\Sigma(\omega_{m},0)-\frac{i}{2}\left(1-\frac{\rvert\Sigma^{q}(\omega_{m})\rvert}{\sqrt{[\Sigma^{q}(\omega_{m})]^{2}+4B}}\right)\epsilon_{\bm{k}},
\end{equation}
where
\begin{equation}
\Sigma(\omega_{m},0)=\frac{\tilde{\Sigma}^{q}(\omega_{n})}{2}+\text{sgn}(\tilde{\Sigma}^{q}(\omega_{n}))\sqrt{\frac{1}{4}[\tilde{\Sigma}^{q}(\omega_{n})]^{2}+B}.
\end{equation}
The first term renormalizes the frequency dependence of the Green's
function,
 while the second
term renormalizes the Fermi velocity into
\begin{equation}
v_{F}^*  =  \frac{1}{2}\left(1+\frac{\rvert\Sigma^{q}(\omega_{m})\rvert}{\sqrt{[\Sigma^{q}(\omega_{m})]^{2}+4B}}\right)v_{F}.
\end{equation}
The renormalized velocity becomes $v_{F}/2$ when $\Sigma^{th} > {\tilde \Sigma}^q$, i.e., when $2\sqrt{B}\gg\Sigma^{q}(\omega_{m})$,
 and differs only slightly from $v_F$ when $\Sigma^{th} < {\tilde \Sigma}^q$.
The crossover between the two regimes is at frequency
\begin{equation}
{\tilde \omega}_{\text{cross}} \sim\frac{B^{3/4}}{\bar{g}^{1/2}}=\bar{g}\left(\frac{T}{\bar{g}}\right)^{3/4}\left(\frac{E_{F}}{\bar{g}}\right)^{3/8}\log^{3/4}\left(\frac{1}{m}\right).
\end{equation}

We next consider the applicability range for Eq. (\ref{bbbb}). Let's  set $\epsilon_k =0$ to avoid unnecessary complications.
 In obtaining (\ref{bbbb}) we assumed that
\begin{equation}
\tilde{\Sigma}_k/v_{F} > m
\label{bbb}
\end{equation}
At $\omega_{m}\ll{\tilde \omega}_{\text{cross}}$, $|\Sigma_k| \approx |\Sigma^{th}| = \sqrt{B}$, and
 the inequality in (\ref{bbb}) sets the condition on $T$:
\beq \label{eq:cond1}
T > \bar{g}\left(\frac{M}{\bar{g}}\right)^{2/3}\left(\frac{E_{F}}{\bar{g}}\right)^{1/2}\frac{1}{\log(1/m)} \sim
\frac{T^{*}}{\log{1/m}}
\eeq
Up to a logarithm, this is $T > T^{*}$, i.e., $T=T^*$ is a sharp boundary between local and non-local forms of $\Sigma^{th}$.  Keeping the logarithm one obtains~\cite{avi} an extended crossover regime. It formally becomes wide at $m \to 0$, but like we said, we assume that
 mode-mode coupling cuts $\log{m}$ at $\log{T}$.  Then the crossover regime is rather narrow.  The upper limit on $T$, at which
 \beq
 T < T_{max} \sim \bar{g}\left(\frac{E_{F}}{\bar{g}}\right)^{1/2}
 \eeq
  is set by the boundary condition on the momentum-independence of the thermal self-energy.

 At $\omega_{m}\ll{\tilde \omega}_{\text{cross}}$,  Eq. (\ref{bbbb}) is valid in the same range of $T$, up to a frequency $\omega \sim {\bar g}$.  We illustrate this in Fig. \ref{fig:cond_nonl}.

\section{Cancellation of $\Sigma^{th}(k)$ in the free energy}
\label{sec:free_app}

In this Appendix, we show explicitly that the thermal self-energy
$\Sigma^{th}$ cancels out in the free energy $F_{el}$, Eq. (\ref{Fe}).
 This holds when $\Sigma^{th}$ is local, and
 when it is non-local and given by (\ref{bbbb}).

\subsection{Case of local self-energy: $\Sigma(k)=\Sigma(\omega_{n})$}

When the self-energy is independent to $\epsilon_{\bm{k}}$, the momentum
integration is straightforward,
\begin{equation}
\sum_{\bm{k}}\log\left(\frac{i\tilde{\Sigma}(\omega)-\epsilon_{\bm{k}}}{\epsilon_{\bm{k}}}\right)=
N_{F}\left(\pi\rvert\tilde{\Sigma}(\omega_{n})\rvert+i\pi\Lambda_q\text{sgn}(\omega_{n})\right),
\end{equation}
\begin{equation}
\sum_{\bm{k}}\frac{\Sigma(\omega)}{\tilde{\Sigma}(\omega)+i\epsilon_{\bm{k}}}=N_{F}\pi\rvert\Sigma(\omega_{n})\rvert.
\end{equation}
Upon summation over $\omega_{n}$ we obtain
\begin{equation}
F_{el}=-2T\sum_{\omega_{n}}\pi N_{F}\left(\rvert\tilde{\Sigma}(\omega_{n})\rvert-\rvert\Sigma(\omega_{n})\rvert\right)\equiv-2\pi TN_{F}\sum_{\omega_{n}}\rvert\omega_{n}\rvert,
\end{equation}
which is equal to the free energy of the non-interacting Fermi gas.
The self-energy cancels out from this expression.

\subsection{Case of non-local $\Sigma^{th} (k)$}

We now show that the cancellation holds even when
  $\Sigma^{th}$ depends on the dispersion $\epsilon_{\bm{k}}$.

   The electronic part of the free energy per volume is
\begin{equation}
F_{el}=-T\sum_{k}\ln\left(\frac{(i\tilde{\Sigma}(k)-\epsilon_{\bm{k}})(i\tilde{\Sigma}(-k)-\epsilon_{-\bm{k}})}{\epsilon_{\bm{k}}^{2}}\right)+2T\sum_{k}\frac{\tilde{\Sigma}(k)-\omega_m }{\tilde{\Sigma}(k)+i\epsilon_{\bm{k}}}.
\end{equation}
 We show below that the non-local $\Sigma^{th} (k)$ actually cancels out in each of two contributions to $F_{el}$.

Substituting $\Sigma^{th}$ from
(\ref{eq:self_nonl})
 into the first term,
we obtain after simple algebra
\begin{align}
(i\tilde{\Sigma}(k)-\epsilon_{\bm{k}})(i\tilde{\Sigma}(-k)-\epsilon_{-\bm{k}}) & =\left(\frac{\rvert\tilde{\Sigma}^{q}(\omega_{m})\rvert\pm i\epsilon_{\bm{k}}}{2}+\sqrt{\frac{\left(\rvert\tilde{\Sigma}^{q}(\omega_{m})\rvert\pm i\epsilon_{\bm{k}}\right)^{2}}{4}+B}\right)\nonumber \\
 & \left(\frac{\rvert\tilde{\Sigma}^{q}(\omega_{m})\rvert\pm i\epsilon_{\bm{k}}}{2}+\sqrt{\frac{\left(\rvert\tilde{\Sigma}^{q}(\omega_{m})\rvert\pm i\epsilon_{\bm{k}}\right)^{2}}{4}+B}\right),
 \label{c}
\end{align}
where $\pm$ refers to $\text{sgn}(\omega_{m})$. Introducing
$\rvert\tilde{\Sigma}^{q}(\omega_{m})\rvert=2\sqrt{B}y$ and $\epsilon_{\bm{k}}=2\sqrt{B}z$,
we re-express (\ref{c}) as
\begin{align}
\frac{(i\tilde{\Sigma}(k)-\epsilon_{\bm{k}})(i\tilde{\Sigma}(-k)-\epsilon_{-\bm{k}})}{\epsilon_{\bm{k}}^{2}} & =\frac{1}{4z^{2}}\left(y\pm iz+\sqrt{\frac{\left(y\pm iz\right)^{2}}{4}+B}\right)^{2},
\end{align}
To obtain the first term in $F_{el}$,  we need to integrate this expression over $\epsilon_k$ (i.e., over $z$) and sum up over Matsubara frequencies.  Combining contribution from positive and  negative $z$, we obtain
\begin{align}
F_{el}^{(1)}= & -T\sum_{k}\ln\left(\frac{(i\tilde{\Sigma}(k)-\epsilon_{\bm{k}})(i\tilde{\Sigma}(-k)-\epsilon_{-\bm{k}})}{\epsilon_{\bm{k}}^{2}}\right)\nonumber \\
= & -4\sqrt{B}N_{F}T\sum_{\omega_{m}}\int_{0}^{\Lambda}dz\ln\left[\frac{1}{4z^{2}}\left(y+iz+
\sqrt{\frac{\left(y+iz\right)^{2}}{4}+B}\right)\left(y-iz+\sqrt{\frac{\left(y-iz\right)^{2}}{4}+B}\right)\right]
\nonumber \\
= & -4\pi\sqrt{B}N_{F}T\sum_{\omega_{m}}y=-2\pi N_{F}T\sum_{\omega_{m}}\rvert\tilde{\Sigma}^{q}(\omega_{m})\rvert.
\end{align}
We see that the result is the same as if $\Sigma^{th}$ was absent.

For the second term in $F_{el}$,  we again use Eq. (\ref{eq:self_nonl}) and express $\Sigma^{th}$ in terms of ${\tilde \Sigma}^q$ and $\epsilon_k$. This yields
\begin{equation}
\frac{\tilde{\Sigma}(k)-\omega_{m}}{\tilde{\Sigma}(k)+i\epsilon_{\bm{k}}}=\frac{\frac{\tilde{\Sigma}^{q}(k)-i\epsilon_{\bm{k}}}{2}+\sqrt{\frac{\left(\tilde{\Sigma}^{q}(k)+i\epsilon_{\bm{k}}\right)^{2}}{4}+B}\text{sgn}(\omega_{m})-\omega_{m}}{\frac{\tilde{\Sigma}^{q}(k)-i\epsilon_{\bm{k}}}{2}+\sqrt{\frac{\left(\tilde{\Sigma}^{q}(k)+i\epsilon_{\bm{k}}\right)^{2}}{4}+B}\text{sgn}(\omega_{m})+i\epsilon_{\bm{k}}}.
\end{equation}
Re-expressing in terms of $y$ and $z$, as did before, we obtain
\begin{align}
 & \sum_{\bm{k}}\frac{\tilde{\Sigma}(k)-\omega_{m}}{\tilde{\Sigma}(k)+i\epsilon_{\bm{k}}}\nonumber \\
= & 2\sqrt{B}N_{F}\int_{-\Lambda_{q}/(2\sqrt{B})}^{\Lambda_{q}/(2\sqrt{B})}dz\frac{y-iz+\sqrt{\left(y+iz\right)^{2}+1}}{y+iz+\sqrt{\left(y+iz\right)^{2}+1}}\nonumber \\
 & -2\rvert\omega_{m}\rvert N_{F}\int_{-\Lambda_{q}/(2\sqrt{B})}^{\Lambda_{q}/(2\sqrt{B})}dz\frac{1}{y+iz+\sqrt{\left(y+iz\right)^{2}+1}}.
\end{align}
This integral is convergent with typical $z=O(1)$. Given that $\Lambda\gg\sqrt{B}$,
the $z-$integration can be extended to infinite limits.  Integrating in infinite limits, we obtain
\begin{align}
\int_{-\infty}^{\infty}dz\frac{y-iz+\sqrt{\left(y+iz\right)^{2}+1}}{y+iz+\sqrt{\left(y+iz\right)^{2}+1}} & =y\int_{-\infty}^{\infty}dt\frac{1-it+\sqrt{\left(1+it\right)^{2}+y^{-2}}}{1+it+\sqrt{\left(1+it\right)^{2}+y^{-2}}}\nonumber \\
 & =\pi y.
\end{align}
\begin{align}
\int_{-\infty}^{\infty}dz\frac{1}{y+iz+\sqrt{\left(y+iz\right)^{2}+1}} & =\int_{-\infty}^{\infty}dt\frac{1}{1+it+\sqrt{\left(1+it\right)^{2}+1}}\\
 & =\pi/2.
\end{align}
Collecting contributions, we find
\begin{align}
F_{el}^{(2)}=2T\sum_{k}\frac{\tilde{\Sigma}(k)-\omega_{m}}{\tilde{\Sigma}(k)+i\epsilon_{\bm{k}}} & =2\pi T N_{F}\left(\rvert\tilde{\Sigma}^{q}(\omega_{m})\rvert-\rvert\omega_{m}\rvert\right)
\end{align}
 as if $\Sigma^{th}$ was absent.
 Combining $F_{el}^{(1)}$
and $F_{el}^{(2)}$,
we obtain
\begin{equation}
F_{el}=-2\pi TN_{F}\sum_{\omega_{m}}\rvert\omega_{m}\rvert,
\end{equation}
which is
the free energy of a non-interacting Fermi gas.
We see that the self-energy cancels out in $F_{el}$  even when $\Sigma^{th}$ depends
on $\epsilon_{\bm{k}}$.

\section{Evaluation of free energy}

\subsection{$\gamma$-model at $\gamma=0^{+}$}

In the purely electronic $\gamma$-model, the free energy is $F_{\gamma}=F_{el}+F_{int}$.
For a generic non-zero $\gamma$, the free energy has been evaluated
in Ref. ~\cite{zhang_22}. Here, we compute the free energy for the special case $\gamma\to0^{+}$,
relevant to the analysis of the antiferromagnetic QCP (see the main
text). The case $\gamma\to0^{+}$ requires special care as the interaction $V(\Omega_m)\propto \log{{\bar g}/|\Omega_m|}$
 For the free
energy, we have in this case
$F_{0+}=F_{free}+ F_{0+,int}$, where  in the notations from the main text
\begin{equation}
F_{0+,int}=\frac{3g^{*}}{8\pi^{3}v_{F}^{2}\beta}S_{0+},
\end{equation}
 $\beta=2v_{x}v_{y}/v_{F}^{2}$, and
\begin{equation}
S_{0+}=(2\pi T)^{2}\sum_{n,n'=-M_{f}}^{M_{f}-1}\text{sgn}(2n+1)\text{sgn}(2n'+1)\log\frac{ \rvert n-n'\rvert 2\pi T}{T_{0}^{**}}.
\end{equation}
The thermal contribution, from $n=n'$, has to be evaluated at a non-zero bosonic mass. This contribution to $F_{0+}$ is linear in $T$ and does not affect the specific heat. Summing over
$n'\neq n$,
 we obtain
\begin{align}
S_{0+} & =4(2\pi T)^{2}\left(2\sum_{n=1}^{M_{f}-1}\log(n!)-\frac{1}{2}\sum_{n=1}^{2M_{f}-1}\log(n!)\right)-4\pi T\Lambda\log{\frac{2\pi T}{T_{0}^{**}}}
\label{d}
\end{align}
Contributions from  $n\sim O(1)$ are of order $\sim T^{2}$
We show that the summation over $n \gg 1$ yields a larger $\sim T^{2}\log(T)$ term.
  To evaluate this contribution,
   we use the asymptotic formula
\begin{equation}
\log(n!)=\left(n+\frac{1}{2}\right)\log(n)-n+\frac{1}{2}\log(2\pi)+\frac{1}{12n}+O(\frac{1}{n^{2}}),\label{eq:exp_f}
\end{equation}
Substituting into (\ref{d}) and using
\bea
&&\sum_{n=1}^{M_{f}-1}\left(n+\frac{1}{2}\right)\log(n)={1\over 2}M_f^2 \log M_f - {1\over 4} M_f^2 - {1\over 2} M_f + O(1) \nonumber \\
&& \sum_{n=1}^{M_{f}-1}\frac{1}{n}=\log\left(M_{f}\right)+O(1), \nonumber \\
&& \sum_{n=1}^{M_{f}-1} n =M_f^2/2 - M_f/2, ~~~\sum_{n=1}^{M_{f}-1} 1 = M_f -1
\eea
and the relation between $M_f$  and the upper theory cutoff $\Lambda$,
 we obtain
\bea
S_{0+} &=& 4 (2\pi T)^2 \left[ -M_f^2 \log 2 + {1\over 2} \log(2\pi) M_f - {1\over 8} \log M_f + O(1) \right] -  2 (2 \pi T)^2 M_f \log \left( {T \over T_0^{**}} \right) \nonumber \\
& = & -\Lambda^{2}\log(16)+4\pi\Lambda T\log \left( \frac{T_{0}^{**}}{2\pi T} \right) - 2\pi^{2}T^{2}\log\left( \frac{\Lambda}{T} \right)+O(T^2).
\eea
Hence
\begin{align}
F_{0+}^{int} & =N_{F}\frac{3g^{*}}{2\pi^{2} \beta E_{F}} \left[ - \Lambda^{2}\log(2) \right. \nonumber \\
&  \left. + \pi \Lambda T\log \left( \frac{T_{0}^{**}}{T} \right) \right. \nonumber \\
&  \left. - {1\over 2} \pi^2 T^{2}\log\left(\frac{\Lambda}{T}\right)+O(T^2) \right].
\end{align}
Differentiating twice with respect to temperature, one obtains the specific
heat
\begin{equation}
C_{0+}^{int}(T)=N_{F}\frac{3g^{*}}{2\pi \beta E_{F}} \left[\Lambda + \pi T \log\left(\frac{\Lambda}{T}\right)+O(T) \right].
\end{equation}
It contains a constant $\propto\Lambda$ and a universal $T\log(1/T)$ term.

For comparison, we evaluate the free energy of the regularized $\gamma$-model, $\bar{F}_{0+} = F_{frre} + \bar{F}_{0+}^{int}$, where
\begin{equation}
\bar{F}_{0+}^{int}=
\frac{3g^{*}}{8\pi^{3}v_{F}^{2}\beta}\bar{S}_{0+},
\end{equation}
and
\begin{equation}
\bar{S}_{0+}=(2\pi T)^{2}\sum_{n,n'=-M_{f}}^{M_{f}-1}\left(\text{sgn}(2n+1)\text{sgn}(2n'+1)-1\right)\log\frac{\rvert n-n'\rvert2\pi T}{T_{0}^{**}}.
\end{equation}
Since the summand is non-zero only when $2n+1$ and $2n'+1$ has opposite
signs, the thermal part with $n=n'$ is avoided. The sum is evaluated
in the same way as for the original $\gamma$-model, and the result is
gives rise to
\begin{align}
\bar{S}_{0+} & =4(2\pi T)^{2}\left(2\sum_{n=0}^{M_{f}-1}\log(n!)-\sum_{n=0}^{2M_{f}-1}\log(n!)\right) - 4(2\pi T)^{2} M_f^2 \log\frac{2\pi T}{T_{0}^{**}}  \nonumber \\
& = 4 (2\pi T)^2 \left[ -M_f^2 \log M_f + \log \left( { e^{3/2} \over 4} \right) M_f^2 -{1\over 12} \log M_f + O (1) \right]  \nonumber \\
&  - 4(2\pi T)^{2} M_f^2 \log\frac{2\pi T}{T_{0}^{**}}   \nonumber \\
 & =4\Lambda^{2}\log\frac{e^{3/2}T_{0}^{**}}{4\Lambda} - \frac{4}{3}\pi^{2}T^{2}\log\left(\frac{\Lambda^2}{2\pi T T_0^{**}}\right)+O\left( T^2 \right).
\end{align}
As expected, the cutoff-dependent $\Lambda T\log(1/T)$ term is removed.
The coefficient of the universal $T^2 \log(1/T)$ term is $2/3$ of that in the original $\gamma$-model. This is the same ratio as for a non-zero $\gamma$ (see the main text).
The interaction part of the free energy is
\begin{align}
\bar{F}_{0+}^{int} & =-N_{F}\frac{3g^{*}}{2\pi^{2} \beta E_{F}}
\left[ \Lambda^{2}\log\left(\frac{4\Lambda}{e^{3/2}T_{0}^{**}}\right) \right. \nonumber \\
 & \left.  + {1\over 3}\pi^2 T^{2}\log\left(\frac{\Lambda^2 }{ 2\pi T T_0^{**}}\right)+O\left( T^2 \right) \right].
\end{align}
Differentiating twice with respect to temperature, we obtain the specific
heat
\begin{equation}
\bar{C}_{0+}^{int}(T)=N_{F}\frac{g^{*}}{\beta E_{F}} T\log\left(\frac{
\Lambda^2}{2\pi T T_0^{**}}\right)+O\left(T \right).
\end{equation}

\subsection{Boson-fermion model}

The free energy of  the underlying boson-fermion model  is given by
$F=F_{free}+F_{bos}$, where
\begin{equation}\label{eq:c15}
F_{bos}=\frac{k}{2} T \sum_{q}\log{\left(- D_{q}^{-1}\right)}
\end{equation}
  and $k$ is
  the number of components of the bosonic fields: $k=1$
     for Ising-nematic and electron-phonon cases, and $k=3$ for an antiferromagnetic QCP.
We presented the results for $F_{bos}$ for the three cases in the main text.  Here
 we show the details of the evaluation of $F_{bos}^{*}$.

\subsubsection{Ising-nematic QCP}

 Subtracting frequency-independent term from $\log\left(- D_{q}^{-1}\right)$ and integrating over the momentum in Eq.~(\ref{eq:c15}) we obtain
\begin{align}
F_{bos} & =\frac{T}{2}\sum_{\Omega_{n}}\int\frac{d^{2}\bm{q}}{4\pi^{2}}\log\left(1+\frac{\alpha\rvert\Omega_{n}\rvert}{q^{3}}\right)=\frac{\alpha^{2/3}}{4\sqrt{3}}T\sum_{\Omega_{n}}\rvert\Omega_{n}\rvert^{2/3}.
\end{align}
The frequency sum over $2M_b+1$ Matsubara frequencies is expressed via the Harmonic number $\sum_{n=1}^{M_{b}}n^{2/3}=H_{-2/3}(M_{b})$. Then
$F_{bos}=\alpha^{2/3}(2\pi T)^{5/3}H_{-2/3}(M_{b})/4\sqrt{3}\pi$.

Using the expansion of Harmonic number at large argument, $H_{-2/3}(M_b) = (3/5) (M_b+1/2)^{5/3} + \zeta(-2/3)+O(1/(M_b+1/2)^{1/3})$, and using the relation between $M_b$ and $\Lambda$, Eq. (\ref{Fe_ex_2}), we obtain
\beq
F_{bos} = {\alpha^{2/3} \over 4\sqrt{3}\pi } \left( {3\over 5} \Lambda^{5/3} + \zeta(-{2\over 3}) (2\pi T)^{5/3}  \right).
\eeq
Differentiating twice over temperature and combining with free-fermion contribution, we obtain $C_{I-N} (T)$, given by
 Eq. (\ref{Fe_14_1}).

\subsubsection{Antiferromagnetic QCP}
\label{sec:afm_app}

For this case, the momentum integral in Eq.~(\ref{eq:c15}) is logarithmically singular and depends on the upper momentum cutoff $\Lambda_q \sim k_F$.  Integrating over $q$, we obtain
\begin{align}
F_{bos} & =\frac{ 3 \alpha T}{8\pi}\sum_{\Omega_{n}}\rvert\Omega_{n}\rvert\log\frac{\Lambda^{2}_q}{\alpha\rvert\Omega_{n}\rvert}\equiv-
\frac{3\alpha}{2}T^{2}\sum_{n=1}^{M_{b}}n\log\frac{nT}{T_{0}},
\label{e}
\end{align}
where $T_{0}\sim \Lambda^2_q/\alpha$. The frequency  sum over $2M_b+1$ Matsubara frequencies is expressed in terms
of the hyperfactorial function $H(x)$ as
\begin{equation}
\sum_{n=1}^{M_{b}}n\log\frac{nT}{T_{0}}=\log\left[H(M_{b})\right]+\frac{M_{b}\left(M_{b}+1\right)}{2}\log\frac{T}{T_{0}}.
\end{equation}
At large $M_{b}\gg1$,  $\log\left(H(M_{b})\right)$ is expanded as
\begin{align}
\log\left(H(M_{b})\right) & =-\frac{1}{4}M_{b}^{2}+\left(\frac{1}{12}+\frac{1}{2}M_{b} (M_b +1) \right)\log(M_{b})\nonumber \\
 & + {\cal O}(1). \label{eq:exp_hf}
\end{align}
Using the relation between
$M_b$ and $\Lambda$,  Eq. (\ref{Fe_ex_2}),
 we obtain after simple algebra
\begin{align}
(2\pi T)^{2}\sum_{n=1}^{M_{b}}n\log\frac{nT}{T_{0}} & =-\frac{1}{4}\Lambda^{2}+\frac{1}{2}\Lambda^{2}\log\frac{\Lambda}{2\pi T_{0}}+\frac{1}{3}\pi^{2}T^{2}\log\frac{T_{0}}{T}+{\cal O}(T^{2}).
\end{align}
Hence
\begin{align}
F_{bos} & =-\frac{3\alpha}{16\pi^{2}}\Lambda^{2}\log\frac{\Lambda}{2\pi T_{0}\sqrt{e}}+\frac{\alpha}{8}T^{2}\log\frac{T}{T_{0}}+{\cal O}(T^{2}).
\end{align}

\subsubsection{QCP of an Einstein phonon}
\label{sec:phonon_app}

Near a QCP at which the dressed Debye frequency vanishes for $q <2k-F$, the dressed phonon propagator takes
the form $D_{q}^{-1}=\Omega_{n}^{2}+\bar{\omega}_{D}^{2}+2\bar{g}^{2}\rvert\Omega_{n}\rvert/(v_{F}q)  (2k_F/\sqrt{4k^2_F -q^2})$,
where $\omega_{D}$ and $\bar{\omega}_{D}=\omega_{D}(1-2\lambda)^{1/2}$
are bare and dressed Debye frequencies, and $\lambda=\bar{g}^{2}/\omega_{D}^{2}$.
Substituting into (\ref{eq:c15}) and treating the Landau damping term as perturbation, we obtain
\begin{align}
F_{bos} & \simeq\frac{T}{2}\sum_{\Omega_{n}}\int\frac{d^{2}\bm{q}}{4\pi^{2}}\log\left(\Omega_{n}^{2}+\bar{\omega}_{D}^{2}\right)+\frac{T}{2}\sum_{\Omega_{n}}\int\frac{d^{2}\bm{q}}{4\pi^{2}}\frac{2\bar{g}^{2}}{v_{F}q}\frac{\rvert\Omega_{n}\rvert}{\Omega_{n}^{2}+\bar{\omega}_{D}^{2}} {2k_F \over \sqrt{4k_F^2 -q^2}}.
\label{eee}
\end{align}
where the integration over $q$ is up to $2k_F$.
The first term is the free energy of a free Einstein phonon with the
dressed Debye frequency $\bar{\omega}_{D}$:
\beq
F_{bos}^{(1)} = 4 N_F E_F T  \left(\log{\bar{\omega}_{D}} +
2 \sum_{n=1}^{M_B} \log ({2\pi T n})
+ \sum_{n=1}^{M_B}
\log\left(1 + \frac{\bar{\omega}_{D}^{2}}{4\pi^2 T^2 n^2}\right)\right)
\label{eeee}
\eeq
Using
\bea
&&\sum_{1}^{M_B}  \log{n} = (M_b +1/2) \log{(M_b +1/2)/e} + \frac{1}{2} \log{2\pi} \nonumber \\
&&\sum_{1}^{M_B}  \log{2\pi T} = (M_b +1/2) \log{2\pi T} - \frac{1}{2} \log{2\pi T}
\eea
 and  the relation between $M_B$ and $\Lambda$, we obtain
 \beq
F_{bos}^{(1)} = 4 N_F E_F \left[\frac{\Lambda}{\pi} \log{\Lambda}{e} +
\sum_{1}^{M_B}  \log\left(1 + \frac{\bar{\omega}_{D}^{2}}{4\pi^2 T^2 n^2}\right)
 - log{T}\right]
\label{eeeee}
\eeq
 The first term is $T$-independent and does not contribute to entropy and specific heat.  In the
  second term, the sum over $m$ converges and the summation can be extended to $M_b = \infty$.
  Evaluating the sum using Euler-Maclauren formula and combining with the last term, we obtain
 \beq
F_{bos}^{(1)} = 4 N_F E_F \left[ {\Lambda \over \pi} \log \left( {\Lambda \over e} \right)+ T \log\left(1-e^{-\bar{\omega}_{D}/T}\right) \right].
\label{eeeeee}
\eeq
We note in passing that the exponential temperature dependence of $F_{bos}^{(1)}$ at the smallest $T$
 implies that all terms in Euler-Maclauren series expansion in $T/{\bar \omega}_D$ vanish, as we explicitly verified.

Carrying out the momentum integration in the second term in (\ref{eee}), we obtain
\begin{align}
F_{bos}^{(2)} & =\pi \bar{g}^{2}N_{F}T\sum_{\Omega_{n}}\frac{\rvert\Omega_{n}\rvert}{\Omega_{n}^{2}+\bar{\omega}_{D}^{2}}=\bar{g}^{2}N_{F} \sum_{n=1}^{M_{b}}\frac{n}{n^{2}+\left(\frac{{\bar \omega}_{D}}{2\pi T}\right)^{2}},
\end{align}
The sum
over  Matsubara frequencies is expressed via di-Gamma functions as
\begin{equation}
\sum_{n=1}^{M_{b}}\frac{n}{n^{2}+\left(\frac{{\bar \omega}_{D}}{2\pi T}\right)^{2}}=\text{Re}\left[\psi\left(1+i\frac{{\bar \omega}_{D}}{2\pi T} +M_{b}\right)-\psi\left(1+i \frac{{\bar \omega}_{D}}{2\pi T}\right)\right].
\end{equation}
Using the asymptotic expression $\psi(z)\simeq\log(z)$ at $\rvert z\rvert\gg1$ and re-expressing
$\log{\Lambda/(2\pi T)}$ as $\log{\Lambda/{\bar \omega}_D} + \log{{\bar \omega}_{D}/(2\pi T)}$  we obtain
\bea
F_{bos} & = &  N_F  \left[4E_F {\Lambda \over \pi} \log \left( {\Lambda \over e} \right)  +
\bar{g}^{2} \log\left( {\Lambda \over \bar{\omega}_D} \right) \right] \nonumber \\
& + & 4  N_{F} E_F T \left[ \log{\left(1-e^{-\bar{\omega}_{D}/T}\right)} +
\lambda_E {\bar{\omega}_D \over 4 T}
 f\left(\frac{\bar{\omega}_{D}}{2\pi T}\right)\right].
\eea
where  the dimensionless function $f(x)$ is
\begin{equation}
f(x)=\log x-\frac{1}{2}\psi(1+ix)-\frac{1}{2}\psi(1-ix).
\end{equation}
This is Eq. (\ref{Fe_22_a}) in the main text.

\section{Phenomenological models that map to the $\gamma$-model with $0<\gamma<1$}
\label{sec:extension_app}

In this Appendix, we  consider a phenomenological extension of the  Ising-nematic model, which maps to the $\gamma$ model with $\gamma =1/3$, to a  family of boson-fermion models that map to the
$\gamma$-model with $0<\gamma<1$. The boson propagator takes the
form
\begin{equation}
D_{q}^{-1}=-\left(q^{2-a}+\frac{\alpha\rvert\Omega\rvert}{q}\right)/ D_0 ,
\end{equation}
where the parameter $a$ is tunable. We assume that the Fermi surface is circular, like in the Ising-nematic case.

To establish the relation with the  $\gamma$-model, we compute the free energy, $F=F_{el}+F_{int}$. As in the Ising-nematic case,
 it can be re-expressed as $F=F_{free}+F_{int}$, where $F_{free}$ is the contribution of free Fermi gas, and $F_{int}$ comes from fermion-boson interaction
\begin{equation}
F_{int}=-g^{2}T^{2}\sum_{m,m^{\prime}}\int\frac{d^{2}\bm{k}d^{2}\bm{k}^{\prime}}{(4\pi^{2})^{2}}\frac{1}{i\tilde{\Sigma}(\omega_{m})-\epsilon_{\bm{k}}}\frac{1}{i\tilde{\Sigma}(\omega_{m}^{\prime})-\epsilon_{\bm{k}^{\prime}}}D_{q},
\end{equation}
where $\bm{q}=\bm{k}-\bm{k}^{\prime}$ by momentum conservation. We assume and then verify that
typical momentum scale in the boson propagator, $\sim\omega^{1/(3-a)}$,
is much larger than the one in the fermion propagator, $\sim\tilde{\Sigma}(\omega)/v_{F}$. In this situation,
 the momentum integration can be factorized as
\begin{align}
F_{int} & =g^{*}T^{2}\sum_{m,m^{\prime}}\int\frac{dk_{\bot}}{2\pi}\frac{1}{i\tilde{\Sigma}(\omega_{m})-v_{F}k_{\bot}}\int\frac{dk_{\bot}^{\prime}}{2\pi}\frac{1}{i\tilde{\Sigma}(\omega_{m}^{\prime})-v_{F}k_{\bot}^{\prime}}\nonumber \\
 & \int\frac{dq_{\parallel}}{2\pi}\frac{1}{|q_{\parallel}|^{2-a}+\frac{\alpha\rvert\omega_{m}-\omega_{m}^{\prime}\rvert}{|q_{\parallel}|}}.
\end{align}
Carrying out the momentum integration, we obtain
\begin{align} \label{eq:d4}
F_{int} & =-\pi^{2}T^{2}N_{F}\bar{g}^{\frac{1-a}{3-a}}\sum_{m,m^{\prime}}\sum_{\bm{k}\bm{k}^{\prime}}\frac{\text{sgn}
(\omega_{m}\omega_{m^{\prime}})}{\rvert\omega_{m}-\omega_{m^{\prime}}\rvert^{\frac{1-a}{3-a}}}.
\end{align}
This is equivalent to the free energy of the $\gamma$-model with $\gamma=(1-a)/(3-a)$ and
  the effective coupling constant
\begin{equation}
\bar{g}=\left(\frac{1}{(3-a)\sin\frac{2\pi}{3-a}}\frac{g^{*}}{2\pi v_{F}\alpha^{\frac{1-a}{3-a}}}\right)^{\frac{3-a}{1-a}}.
\end{equation}
The effective $\gamma$ changes continuously from $0$ to $1$ when $a$ is changes between $1$ to $-\infty$. For all these  $a$, the coupling constant $\bar{g}^{\gamma}$ remains positive-defined.
The sum in Eq.~(\ref{eq:d4}) has been evaluated in the main text. It contains $\Lambda-$dependent terms and the universal term of order $T^{(5-a)/(3-a)}$.
In the regularized $\gamma$-model, $\Lambda-$dependent terms cancel out. The free energy is
\begin{align}
\bar{F}_{\gamma} & =F_{free}+\frac{2}{4(3-a)\sin\frac{2\pi}{3-a}}\zeta\left(-\frac{2}{3-a}\right)\left(2\pi\alpha\right)^{\frac{2}{3-a}}T^{\frac{5-a}{3-a}}.
\end{align}
The full free energy of the model includes the contribution from bosons.
\begin{equation}
F_{full} = F_{full}(T=0) - {\pi^2 \over 3} N_F T^2 + \frac{1}{4\sin\frac{2\pi}{3-a}}\zeta\left(-\frac{2}{3-a}\right)\left(2\pi\alpha\right)^{\frac{2}{3-a}}T^{\frac{5-a}{3-a}},
\end{equation}
where $F_{full}(T=0)$ comes from the zero-temperature quantum fluctuations and depends on cutoff $\Lambda$.
Comparing the $T$-dependent terms in $F_{full}$ and $\bar{F}_{\gamma}$, we see that they have the same form, but the
 prefactors for the $T^{(5-a)/(3-a)}$ term differ by $2/(3-a)$.
The prefactors agree at $a =1+0$, when $\gamma = 0+$, as we also found in the explicit analysis of the $\gamma = 0+$ model in the main text.

\bibliography{free_energy_1.bib}

\end{document}